\documentclass[sn-mathphys]{sn-jnl}

\usepackage{comment}

\usepackage{amsmath}
\usepackage{amssymb}
\usepackage{natbib}
\usepackage{graphicx}
\usepackage{subfig}
\usepackage{float}
\newcommand{\be}{\begin{equation}}
\newcommand{\ee}{\end{equation}}
\newcommand{\bea}{\begin{eqnarray}}
\newcommand{\eea}{\end{eqnarray}}

\begin{document}

\title{Generalized Hilltop Inflation}

\author[1]{\fnm{Helena Grete} \sur{Lillepalu}}

\author*[1]{\fnm{Antonio} \sur{Racioppi}}\email{antonio.racioppi@kbfi.ee}

\affil*[1]{\orgname{National Institute of Chemical Physics and Biophysics}, \orgaddress{\street{R\"avala 10}, \city{Tallinn}, \postcode{10143}, \country{Estonia}}}

\abstract{We study a generalized version of hilltop inflation where the standard hilltop potential has been raised to a power and we allow fractional numbers for both the original hilltop power ($m$) and the overall exponent ($n$). In the parameter space studied, agreement with the latest experimental constraints favors high values for $m$ and low values for $n$. Finally, we also find that in all the configurations studied, the inflationary scale always sits around the grand unification scale.}

\keywords{Inflation, hilltop}

\maketitle

\section{Introduction} \label{sec:Introdution}

Past and recent observations of the cosmic microwave background radiation (CMB) support the overall homogeneity and flatness of the Universe  at large distances. The most natural and simplest explanation of such properties is an accelerated expansion of the Universe during its very early stages \cite{Starobinsky:1980te,Guth:1980zm,Linde:1981mu,Albrecht:1982wi}. This inflationary phase can also produce and preserve the primordial inhomogeneities which generated the subsequent large-scale structure that we observe. The simplest way to formulate inflation is by adding a scalar field, the inflaton, to the Einstein-Hilbert action: the scalar field's energy density induces the near-exponential expansion (e.g. \cite{encyclopedia} and refs. therein).

The latest combination from the Planck, BICEP/Keck Array and BAO data \cite{BICEP:2021xfz} has reduced even more the available parameter space, strongly favouring concave inflationary potentials over convex ones. One of the simplest ways to write concave potentials is via hilltop models of inflation (e.g. \cite{encyclopedia,hilltop,Kallosh:2019jnl,Lin:2019fdk,Kohri:2007gq,Dimopoulos:2020kol,German:2020rpn,Hoffmann:2021vty} and refs. therein). The most attractive feature of these models is that any potential with a concave shape can always be approximated with a hilltop potential (at least in some regions of the parameter space). On the other hand, the biggest limitation is that they can only provide an effective description since the hilltop potentials are clearly unbounded from below. The simplest way to solve such an issue is to square the hilltop potential (e.g. \cite{Kallosh:2019jnl,Hoffmann:2021vty}) generating a sombrero-hat potential, like done with the symmetry-breaking inflaton model (e.g. \cite{encyclopedia, OLIVE1990307}). 

The aim of our work is to study a generalized version of the hilltop model. For the sake of curiosity and a better understanding of the effective inflationary behaviour, instead of using an overall power of two (as done in the aforementioned symmetry-breaking models), we consider a generic overall exponent. Inevitably, the price to pay will be falling back into an effective description. On the other hand, we will gain valuable information about a parameter space that, according to our knowledge, has not been explored yet.

This article is organized as follows. In Section \ref{sec:GHI} we revisit the basics of hilltop inflation and we propose our generalized version of it. In Section \ref{sec:Results} we present our numerical results and in Section \ref{sec:Conclusions} we draw our conclusions. Finally, in Appendix \ref{sec:appendix} we show an example of how to get an inflaton potential involving fractional powers from a general scalar-tensor theory. 

\section{A novel model for hilltop inflation} \label{sec:GHI}
\subsection{Hilltop inflation} \label{subsec:HI}
\begin{figure}[t]
    \centering
     \includegraphics[scale=0.6]{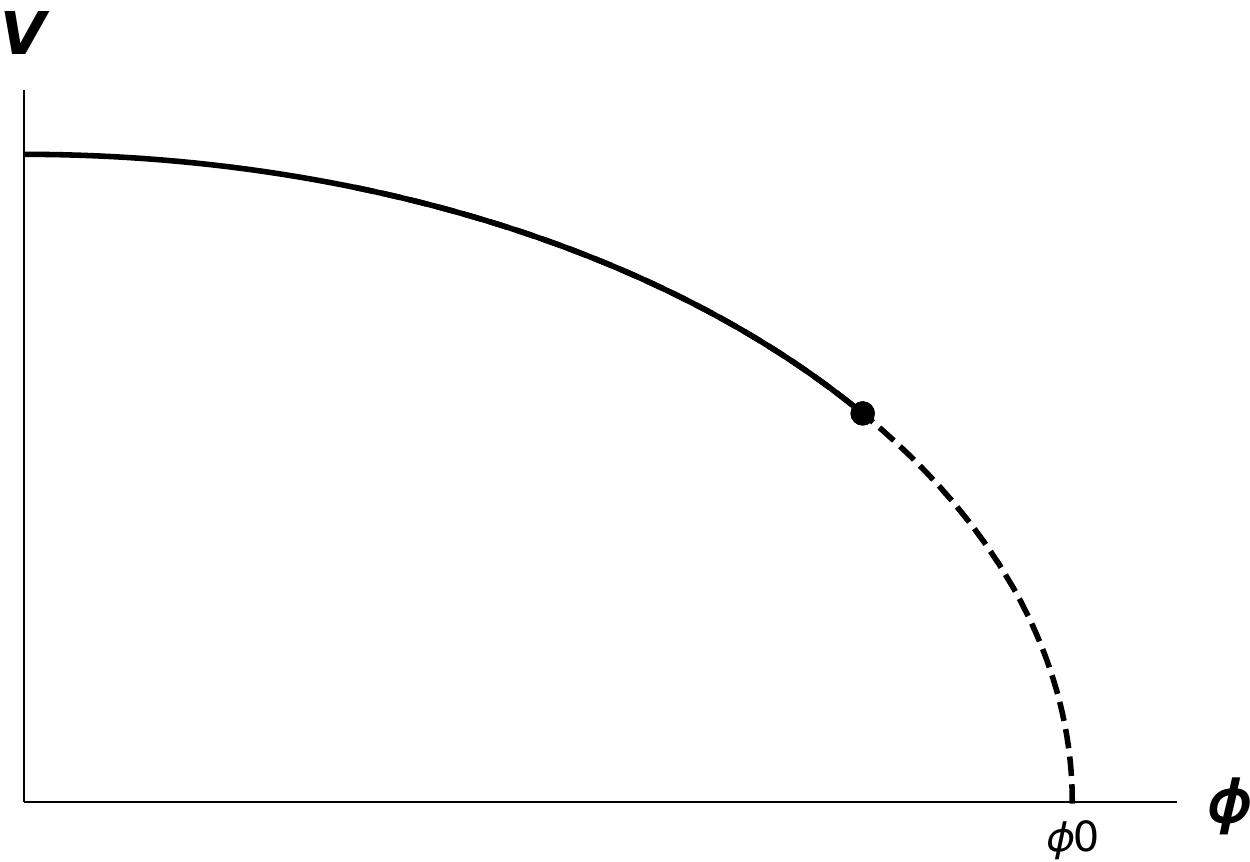}
       \caption{Indicative plot for $V(\phi)_\text{HI}$ vs. $\phi$. The continuous line represents $V(\phi)_\text{HI}$ during inflation, the dot represents the end of inflation and the dashed line represents $V(\phi)_\text{HI}$ after the end of inflation, where the hilltop expansion of the potential loses its validity.
    }
    \label{fig:V-plot}%
\end{figure}

In this Subsection, we remind the reader of the main features of hilltop inflation  (e.g. \cite{encyclopedia,hilltop,Kohri:2007gq,Dimopoulos:2020kol,German:2020rpn,Hoffmann:2021vty,Kallosh:2019jnl,Lin:2019fdk} and refs. therein). The name ``hilltop" comes from the idea that the inflaton potential has the shape of a hill and the inflaton is rolling down from the top of such a hill. Therefore, the defining characteristic of these models is inflation happening near a maximum of the inflaton potential. The hilltop model is defined by an inflaton potential in the form of 
\begin{equation}
\label{eqn:original-hilltop}
    V(\phi)_{\rm HI} = V_0 \bigg[ 1 - \bigg( \frac{\phi}{\phi_0} \bigg)^m \bigg] + \dots,
\end{equation}
where the dots stand for additional stabilizing terms assumed to be irrelevant in the inflationary regime but needed for keeping the potential bounded from below and $V_0$, $\phi_0$ and $m$ are all positive constants. Inflation takes place in the positive quadrant $(\phi \geq 0)$ and the potential is chosen so that the maximum takes place at $\phi=0$. In order to ensure that, the first and second derivatives of the potential in eq. (\ref{eqn:original-hilltop}) 
need to be constrained so that
\begin{eqnarray}
    &&V'(0)_{\rm HI} = 0, \label{eq:V0:HI}\\
\label{eqn:V-second-der-less-zero:HI}
    &&V''(0)_{\rm HI} < 0. 
\end{eqnarray}
Assuming all the parameters positive, eqs. \eqref{eq:V0:HI} and (\ref{eqn:V-second-der-less-zero:HI}) hold only if $m>1$. 
An indicative plot of a generic hilltop potential is given in Fig. \ref{fig:V-plot}. In continuous line, we have $V(\phi)_\text{HI}$ in the inflationary regime and in dashed $V(\phi)_\text{HI}$ after the end of inflation, marked by the dot, where the validity of the expansion in eq. \eqref{eqn:original-hilltop} is soon lost. It is easy to prove that the asymptotic limit of hilltop models is linear inflation regardless of the value of $m$. First of all, taking into account that inflation always takes place for $\phi<\phi_0$, we perform a convenient change of variable
\begin{equation}
    \phi \to \phi_0(1-x)
\end{equation}
so that the inflaton potential \eqref{eqn:original-hilltop} becomes 
\begin{equation}
    V(\phi(x))_\text{HI} = V(x)_\text{HI} = V_0(1 - (1 - x)^m).
\end{equation}
It can be proven that when $\phi_0$ is much larger than the reduced Planck mass $M_\text{P}$, then $x \ll 1$, but never exactly 0. It is important that $x \neq 0$ because otherwise, the slow-roll approximation would be invalid. However, because $x \ll 1$, it is possible to take a Taylor series at the first order in $x$, which will simplify the potential to
\begin{equation}
    V(x)_\text{HI} \simeq V_0 m x, \label{eq:linear:limit}
\end{equation}
which is a linear potential in $x$. Therefore the asymptotic limits of the inflationary predictions are the ones of linear inflation, in agreement with the previous studies on the topic (e.g. \cite{hilltop,Kohri:2007gq,Dimopoulos:2020kol,German:2020rpn,Hoffmann:2021vty,Kallosh:2019jnl,Lin:2019fdk} and refs. therein).

\subsection{Generalized hilltop inflation}
The most common and popular way to stabilize the potential in eq. \eqref{eqn:original-hilltop} is to convert it into a sombrero-hat potential (e.g. \cite{Hoffmann:2021vty,Kallosh:2019jnl,OLIVE1990307})
\begin{equation}
\label{eqn:sombreo}
    V(\phi) = V_0 \bigg[1-\bigg( \frac{\phi}{\phi_0}\bigg)^m \bigg]^2 . 
\end{equation}
In this article, for the sake of curiosity and research, we give up the attempt of stabilization and relax the overall power that is currently 2 and consider it as a new parameter $n$, obtaining the following potential 
\begin{equation}
\label{eqn:potential}
    V(\phi) = V_0 \bigg[1-\bigg( \frac{\phi}{\phi_0}\bigg)^m \bigg]^n, 
\end{equation}
where $m$, $n$, $V_0$, $\phi_0>0$ are free parameters and only positive values for $\phi$ are considered as well. The potential in eq. \eqref{eqn:potential} is labelled as the generalized hilltop inflation (GHI) model. Again first, we need to check that the potential possesses a maximum in $\phi=0$:
\begin{eqnarray}
    &&V'(0) = 0, \label{eq:V0}\\
\label{eqn:V-second-der-less-zero}
    &&V''(0) < 0. 
\end{eqnarray}
 Therefore, the constraints \eqref{eq:V0} and \eqref{eqn:V-second-der-less-zero} imply again $m>1$ (as in the usual hilltop model) while the only constraint on $n$ is to be positive. To conclude, the indicative plot for $V(\phi)$ is as well given in Fig. \ref{fig:V-plot}.

\subsection{Inflationary computations}
\label{section:inflation-recipe}
A summary of the slow-roll formalism can be found in \cite{encyclopedia}. Here we will just give the relevant equations. Inflation takes place when the first and second slow-roll parameters
\begin{eqnarray}
\label{eqn:ev_formula}
   \epsilon_V (\phi) &=& \frac{M_\text{P}^2}{2} \bigg( \frac{V'(\phi)}{V(\phi)}\bigg)^2\, , \\
\label{eqn:nv_formula}
    \eta_V (\phi) &=& M_\text{P}^2\frac{V''(\phi)}{V(\phi)} \, ,
\end{eqnarray}
satisfy $\epsilon_V,\eta_V \ll 1$.
The number of e-folds of expansion of the Universe is given by
\be
N_* = \frac{1}{M_{\rm P}^2} \int_{\phi_{\text{end}}}^{\phi_N} {\rm d}\phi \, \frac{V(\phi)}{V'(\phi)} ,
\label{eq:Ne}
\ee
where the field value at the end of inflation is given by $\epsilon_V(\phi_{\text{end}}) = 1 $, while the field value $\phi_N$ at the time a given scale left the horizon is given by the corresponding $N_*$. The two main observables are the spectral index $n_\text{s}$ and the tensor-to-scalar ratio $r$:
\bea
n_\text{s} &=& 1+2\eta_V(\phi_N)-6\epsilon_V(\phi_N) \, , \\
r &=& 16\epsilon_V(\phi_N) \,  .
\eea
Finally, the amplitude of the scalar power spectrum, 
\be
 A_\text{s} = \frac{1}{24 \pi^2 M_\text{P}^4}\frac{V(\phi_N)}{\epsilon_V(\phi_N)} \, ,
 \label{eq:As:th}
\ee
has to satisfy \cite{Planck2018:inflation}
\be
\label{eq:As:constraint}
\ln \left(10^{10} A_\text{s} \right) = 3.044 \pm 0.014   \, .
\ee
The equations for $\epsilon_V$ and $\eta_V$ turn out be
\begin{eqnarray}
\label{eqn:eVGeneral}
   &&\hspace{-1.3cm}\epsilon_V = \frac{m^2n^2M_\text{P}^2}{2}  \frac{\phi^{2m - 2}}{\big( \phi_0^m - \phi^m \big)^2} \, ,\\
\label{eq:eta:V}
 &&\hspace{-1.3cm}\eta_V = m n M_\text{P}^2\, \frac{\Big(\frac{\phi}{\phi_0}\Big)^m\Big(1-m+(mn-1)\Big(\frac{\phi}{\phi_0}\Big)^m\Big)}{\phi^2\Big( \Big(\frac{\phi}{\phi_0}\Big)^m -1\Big)^2} \, .
\end{eqnarray}
From eq. (\ref{eqn:eVGeneral}) it is clear that 
\begin{equation}
    \lim_{\phi  \to \, 0} \epsilon_V = 0,
\end{equation}
\begin{equation}
\label{eqn:ev-inf-at-phi0}
    \lim_{\phi \to \, \phi_0} \epsilon_V = \infty \, .
\end{equation}
The number of $e$-folds of inflation can be computed as
\begin{equation}
N_* = \left\{ 
 \begin{array}{ll}
 \left. \frac{\phi^2}{2 m n M_\text{P}^2} \left( 1 + \frac{2\left(\frac{\phi}{\phi_0}\right)^{-m}}{m - 2} \right) \right\vert^{\phi_N}_{\phi_{\text{end}}} & \text{for } m \neq 2 \\
 \left. \frac{\phi^2 - 2\phi_0^2\ln(\phi)}{4n M_\text{P}^2}\right\vert^{\phi_N}_{\phi_{\text{end}}} & \text{for } m = 2 \, .
\end{array}
\right.
\end{equation}
Then, the tensor-to-scalar ratio and the scalar spectral index $n_\text{s}$ are respectively
\begin{eqnarray}
\label{eqn:r-detailed}
  &&\hspace{-1.3cm}  r = 8m^2n^2M_\text{P}^2 \frac{\phi_N^{2m-2}}{(\phi_0^m-\phi_N^m)^2} \, ,\\
 &&\hspace{-1.3cm}   n_\text{s} = 1- mnM_\text{P}^2\, \frac{\Big(\frac{ \phi_N}{\phi_0} \Big)^m\Big(2m-2+\Big(mn+2\Big)\Big(\frac{ \phi_N}{\phi_0} \Big)^m\Big)}{ \phi_N^2 \Big(\Big(\frac{ \phi_N}{\phi_0} \Big)^m -1 \Big)^2}. \label{eqn:ns-detailed}
\end{eqnarray}
Finally, the amplitude of the scalar power spectrum is
\begin{equation}
    A_\text{s} = \frac{V_0}{12m^2n^2\pi^2M_\text{P}^6} \,  \phi_N^2 \bigg(\frac{\phi_0}{ \phi_N}\bigg)^{2m}\bigg(1- \bigg(\frac{ \phi_N}{\phi_0}\bigg)^m \bigg)^{n+2}.
\end{equation}

\section{Results} \label{sec:Results}
We performed inflationary computations dealing with the various parameters in the following way. 
 First of all, we considered convenient to write the two energy scales, $V_0$ and $\phi_0$, in terms of Planck masses. Therefore, the normalisation of the potential $V_0$ was parametrized as
\begin{equation}
    V_0 = \left( \lambda M_\text{P} \right)^4. \label{eq:V0lambda}
\end{equation}
and $\lambda$ is fixed in order to satisfy the constraint \eqref{eq:As:constraint}.
%
Then, we parametrized $\phi_0$ as
\begin{equation}
\label{eqn:delta-def}
    \phi_0 = \delta\ M_\text{P} \, ,
\end{equation}
and varied $\delta$ between\footnote{Compatibility with latest combination of Planck, BICEP/Keck and BAO data \cite{BICEP:2021xfz} forces us to consider transPlanckian $\phi_0$'s. This will result in tranPlanckian inflaton field fluctuations, implying a breaking of the Lyth bound \cite{Lyth:1996im} and the eventual appearance of higher order operators that could potentially jeopardize the shape of the inflaton potential. Therefore the completion of the theory would need to be so that our effective model is protected against those operators. However, the construction of such a completion is beyond the scope of our paper, therefore, for the purposes of our analysis, we just assume its existence. On the other hand, the Lyth bound and its meaning are strongly debated. For instance, it has been proposed that transPlanckian field values are still acceptable when the energy density remains subPlanckian \cite{Linde:2004kg}, which is exactly what happens in our case (at least in the region in agreement with data): the constraint on the amplitude of scalar perturbations \cite{Planck2018:inflation} will imply $\lambda<1$ (see Figs. (d) from \ref{fig:m1.5} till \ref{fig:m4_filt}). We leave the reader free to choose the preferred approach among the two presented.   
}
$1$ and $1000$. 
Smaller $\delta$'s were not considered because otherwise, the slow-roll approximation loses validity. Larger $\delta$'s were not considered because at large $\delta$ the values for $r$ and $n_\text{s}$ approach their asymptotic limits and do not change sensibly anymore. 
Finally, for $m$ and $n$ all combinations of $m\in\{\frac{3}{2}, 2, 3, 4\}$ and $n\in\{\frac{1}{4}, \frac{1}{2}, 1, 2, 3, 4\}$ were used. Later the results will show that such an ensemble is big enough to provide relevant information. The use of fractional powers might seem unnatural. However, they can be easily realized in the larger framework of scalar-tensor theories (see e.g. \cite{Jarv:2016sow,Jarv:2020qqm} and refs. therein). For more details see the Appendix \ref{sec:appendix}. The models where $n = 1$ correspond to standard hilltop inflation  (e.g. \cite{hilltop,Kohri:2007gq,Dimopoulos:2020kol,German:2020rpn,Hoffmann:2021vty,Kallosh:2019jnl,Lin:2019fdk} and refs. therein), described by the potential in equation (\ref{eqn:original-hilltop}), while the configuration with $m=2$ and $n=2$ is equivalent to the symmetry-breaking model  (e.g. \cite{OLIVE1990307}).

The plots in Figs. \ref{fig:m1.5},  \ref{fig:m2}, \ref{fig:m3}, \ref{fig:m4} show the results respectively when  $m = \frac{3}{2}, 2, 3, 4$. For all plots, different colours show different values of $n$. Purple band indicates $n = \frac{1}{4}$, cyan  $n = \frac{1}{2}$, green  $n = 1$, blue $n = 2$, yellow $n = 3$ and red $n = 4$.
In the background in grey, there are the latest $1\sigma$ and $2 \sigma$ BICEP/Keck constraints from 2021 \cite{BICEP:2021xfz}.  
The constraints are for $r$ vs. $n_\text{s}$, therefore for the $r$ vs. $\delta$ plot and  $\delta$ vs. $n_\text{s}$ plot, the corresponding projection of the constraints is displayed respectively only for $r$ or $n_\text{s}$. Analogously, the plots in Fig. \ref{fig:m1.5_filt},  \ref{fig:m2_filt}, \ref{fig:m3_filt}, \ref{fig:m4_filt} show respectively the previous results only for the points that are inside the experimental constraints for $r$ and $n_\text{s}$ \cite{BICEP:2021xfz}.

\subsection{General behaviour}
\label{section:general-behaviour}

The general behaviour on the $r$ vs. $n_\text{s}$ plot is the same, regardless of the values of $m$ and $n$. The bands are always almost horizontal at lower $n_\text{s}$ and then start to curve up to higher $r$ for higher $n_\text{s}$. Increases in $m$ and $n$ both increase the thickness of the bands and an increase in $n$ also raises the bands higher in terms of $r$.

The general behaviour on the $r$ vs. $\delta$ plot is the same, regardless of the values of $n$ and $m$. When $\delta$ is small, the values of $r$ are close to 0. As $\delta \to \infty$ the values of $r$ approach a constant value and no longer increase as $\delta$ increases. This constant value depends only on $N_*$ and $n$, but not $m$.

The general behaviour on the $\delta$ vs. $n_\text{s}$ plot is the same, regardless of the values of $n$ and $m$. When $\delta$ is small, the values of $n_\text{s}$ are relatively small. When $\delta \to \infty$, then $n_\text{s}$ approaches a constant value, which once again depends only on $N_\text{e}$ and $n$, but not $m$. For higher values of $n$, more of a ``bend" can be seen on the plot, meaning that the highest values of $n_\text{s}$ are not reached when $\delta \to \infty$, but rather at around $\delta = 50$.

The general behaviour on the $\lambda$ vs. $\delta$ plot is the same, regardless of the values of $n$ and $m$. When $\delta$ is small, then $\lambda$ is relatively small as well. In all different configurations, $\lambda \lesssim 0.01$ (i.e. inflation happening around or before the Grand Unified Theory (GUT) scale) seems achievable.  Then as $\delta$ increases, $\lambda$ increases as well until it follows its asymptotic behaviour and becomes proportional to $\delta$ raised to some power (Figs.  \ref{fig:m1.5},  \ref{fig:m2}, \ref{fig:m3}, \ref{fig:m4} show a linear dependence between $\lambda$ and $\delta$ plotted in a logarithmic scale) and such power increases with $n$ increasing. 
The asymptotic behaviour for $\delta \to \infty$ has the following explanation. It is well known (see eq. \eqref{eq:linear:limit}) that standard hilltop inflation behaves asymptotically as linear inflation regardless of the actual value of $m$. Therefore it is straightforward to derive that in GHI the asymptotic behaviour is monomial inflation. 
Repeating the same steps as in Subsection \ref{subsec:HI}, we obtain
\begin{equation}
    V(\phi(x)) = V(x) = V_0(1 - (1 - x)^m)^n.
\end{equation}
which can be approximated for $x \ll 1$ as
\begin{equation}
    V(x) \simeq V_0 m^n x^n. \label{eq:monomial:limit}
\end{equation}
From this, it can be shown that $r$ and $n_\text{s}$ approach the corresponding asymptotic values of monomial inflation, when $\delta \to \infty$. These values however do not depend on $m$, because $x^n$ does not depend on $m$ and $m^n$ gets reduced away when calculating $\epsilon_V$ and $\eta_V$ (and therefore $r$ and $n_\text{s}$). Moreover, it can be easily checked that $x \propto \frac{1}{\delta }$ at the inflationary scale. Therefore, by applying the constraint \eqref{eq:As:th} we obtain $\lambda \propto \delta^\frac{n}{4}$, explaining the results in  Figs.  \ref{fig:m1.5},  \ref{fig:m2}, \ref{fig:m3}, \ref{fig:m4}.
 The approximation in the $\delta \to \infty$ limit is in agreement with results of the previous studies for $n=2$ (e.g. \cite{Hoffmann:2021vty,Kallosh:2019jnl,OLIVE1990307} and refs. therein).

Before concluding and moving to the more detailed analysis, we stress that all our numerical findings are in agreement with the previous studies concerning the configurations $(m,1)$ and $(m,2)$ (e.g. \cite{hilltop,Kohri:2007gq,Dimopoulos:2020kol,German:2020rpn,Hoffmann:2021vty,Kallosh:2019jnl,Lin:2019fdk} and refs. therein). The only eventual difference comes from the compatibility with the experimental constraints \cite{BICEP:2021xfz} used in our analysis but not available yet in the previous ones.

\subsection{Model with $m=\frac{3}{2}$}
\subsubsection{Unfiltered results}
\begin{figure}[t]
    \centering
     \subfloat[]{\includegraphics[width=0.5\textwidth]{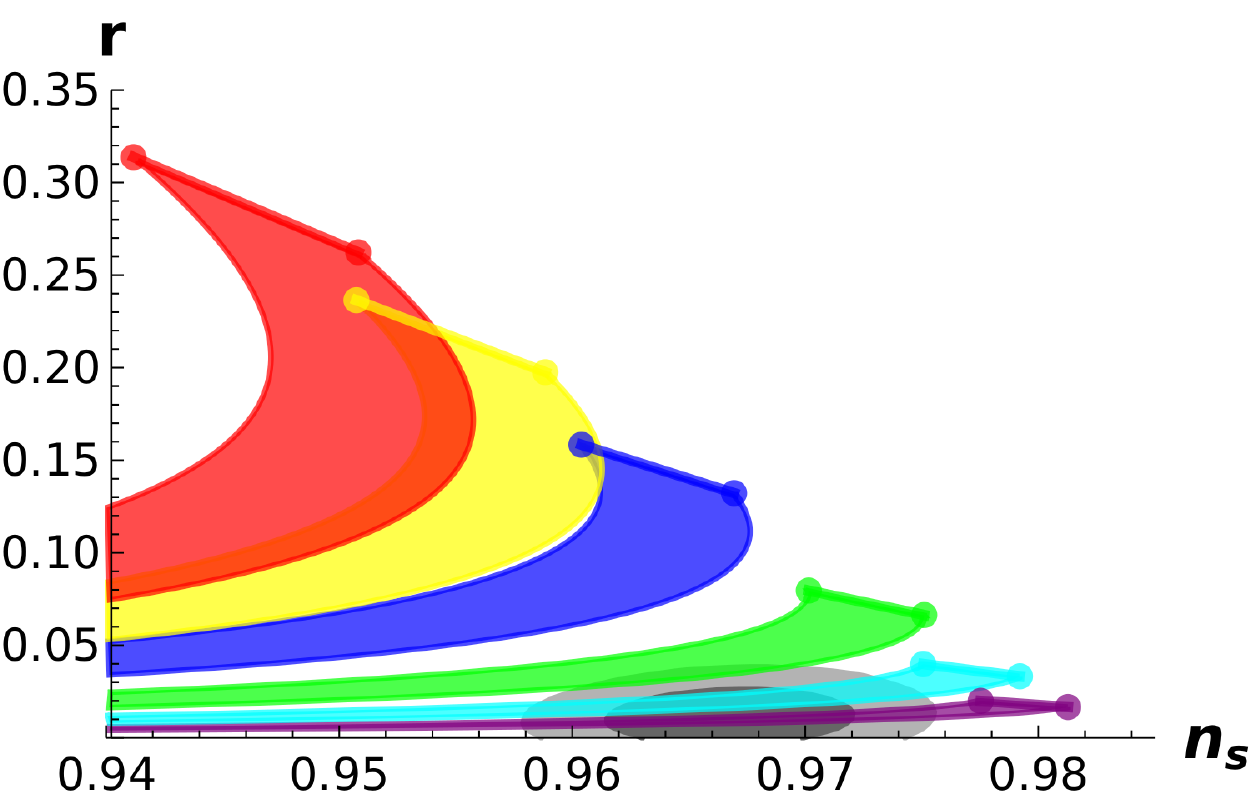}}%
    \subfloat[]{\includegraphics[width=0.5\textwidth]{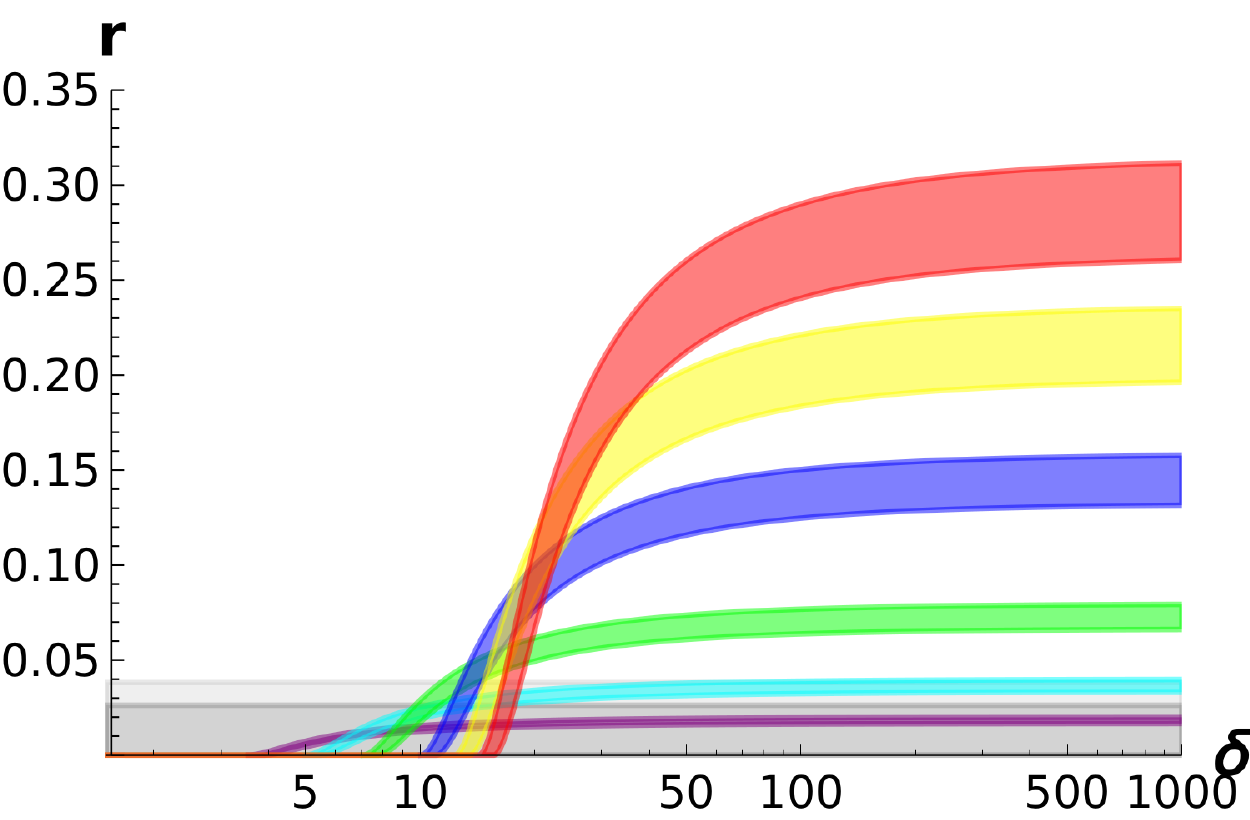}}%

    \subfloat[]{\includegraphics[width=0.5\textwidth]{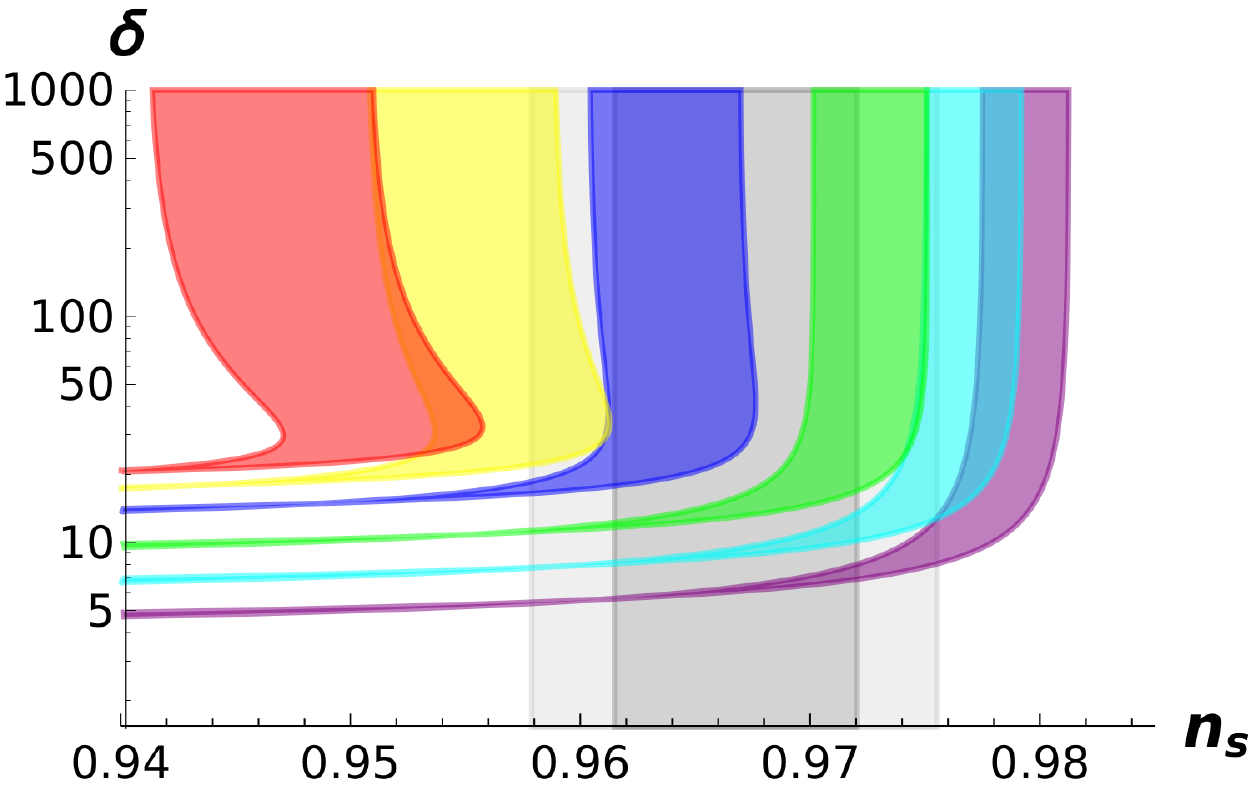}}%
    \subfloat[]{\includegraphics[width=0.5\textwidth]{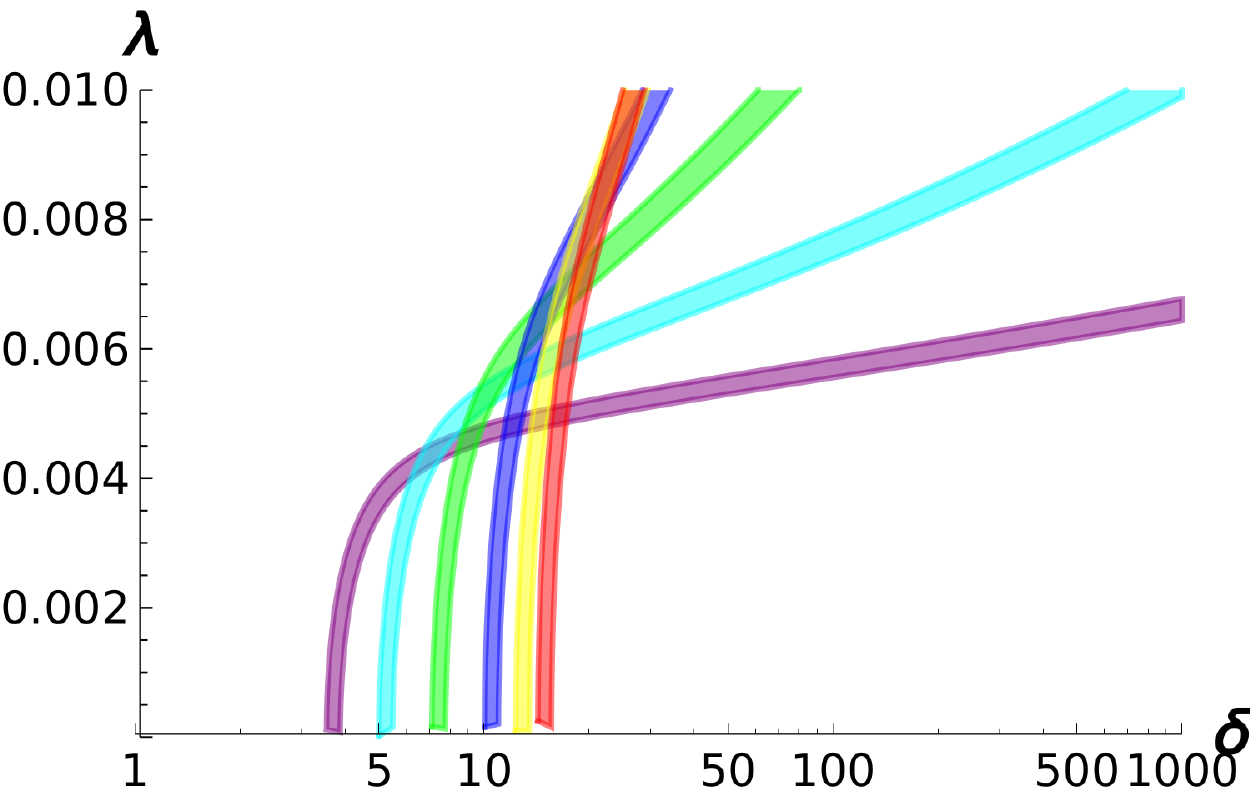}}%

    \caption{$r$ vs. $n_\text{s}$ (a),  $r$ vs. $\delta$ (b),  $\delta$ vs. $n_\text{s}$ (c),  $\lambda$ vs. $\delta$ (d). For these plots $m = \frac{3}{2}$ and $N_* \in [50 , 60]$. Red bands represent the model with $n = 4$, yellow $n = 3$, blue $n = 2$, green $n = 1$, cyan $n = \frac{1}{2}$ and purple $n = \frac{1}{4}$. In a continuous line, between bullets, with the same colour code, there are the corresponding monomial inflation limits. With grey are the 1,2$\sigma$ allowed regions coming  from  the latest combination of Planck, BICEP/Keck and BAO data \cite{BICEP:2021xfz}.} %
    \label{fig:m1.5}%
\end{figure}

The plots in Fig. \ref{fig:m1.5} give a general overview of the behaviour between different parameters when $m = \frac{3}{2}$.

Fig. \ref{fig:m1.5}(a) shows $r$ vs. $n_\text{s}$, for $m = \frac{3}{2}$. Each value of $n$ gives a different band on the plot. For all bands, $r$ is nearly constant at smaller values of $n_\text{s}$ and then it increases faster with $n_\text{s}$ increasing. The higher the value of $n$, the higher the values of $r$ are and the upward trend starts at smaller values of $n_\text{s}$. For $n$ higher than 2, at the highest values of $r$, the values of $n_\text{s}$ start to decrease as well. The higher edge of the band is always the line where $N_* = 50$ and the lower edge of the line where $N_* = 60$. Therefore the smaller the value of $N_*$, the higher the values of $r$ and the upward trend in values of $r$ compared to $n_\text{s}$ starts at smaller $n_\text{s}$ as well. Also, the higher the values of $n$, the wider the bands are.

In Fig. \ref{fig:m1.5}(b) $r$ vs.  $\delta$ is plotted, for $m = \frac{3}{2}$. For each value of $n$, with the increase in $\delta$, the values for $r$ are increasing quickly at first and then at an increasingly slower pace. Also, as $n$ gets larger, the maximum value that $r$ reaches gets larger as well, but higher values of $\delta$ are needed to reach that maximum value. This behaviour is expected based on the asymptotic limit of the inflaton potential given in eq. \eqref{eq:monomial:limit}.

In Fig. \ref{fig:m1.5}(c) $n_\text{s}$ vs.  $\delta$ is plotted, when $m = \frac{3}{2}$. For each value of $n$, with the increase in $\delta$, the values for $n_\text{s}$ are increasing very quickly at first, but then almost not at all. For $n$ = 2 and higher the values for $n_\text{s}$ start to slightly decrease at around $\delta = 30$, with higher values of $n$ giving higher decreases in $n_\text{s}$. 
The higher the values of $n$, the higher the values of $n_\text{s}$ eventually reached. This behaviour is again expected from the asymptotic limit of the inflaton potential in eq. \eqref{eq:monomial:limit}.

In Fig. \ref{fig:m1.5}(d) $\lambda$ vs.  $\delta$ is plotted, when $m = \frac{3}{2}$. For each value of $n$,  with the increase in $\delta$, the values for $\lambda$ increase as well. For fixed smaller values of $\delta$, $\lambda$ decreases with $n$. Then at fixed larger $\delta$ values the situation is reversed with $\lambda$ increasing with $n$.  When $\delta \lesssim 25$, $\lambda \lesssim 0.01$ (or even much smaller) for all the considered cases. This means that the inflationary scale is around or before the GUT scale.


\begin{figure}[t]
    \centering
     \subfloat[]{\includegraphics[width=0.5\textwidth]{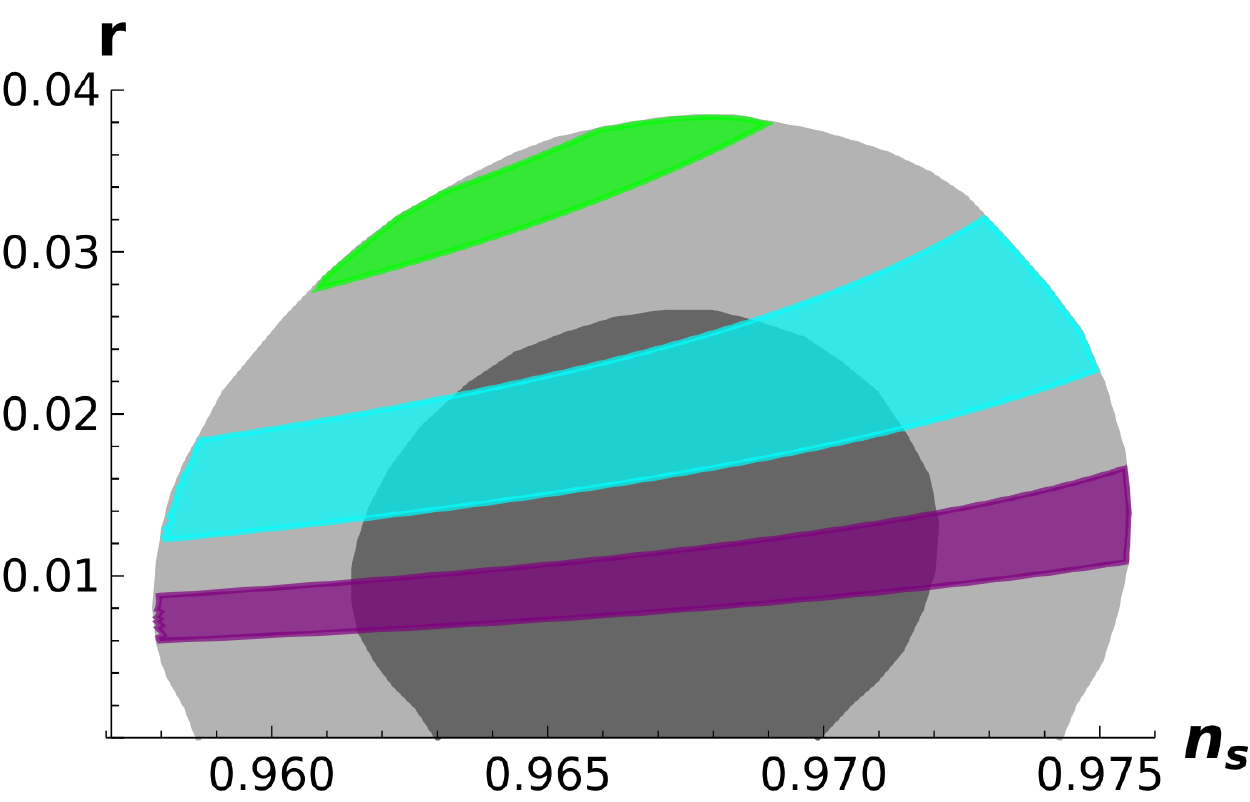}}%
    \subfloat[]{\includegraphics[width=0.5\textwidth]{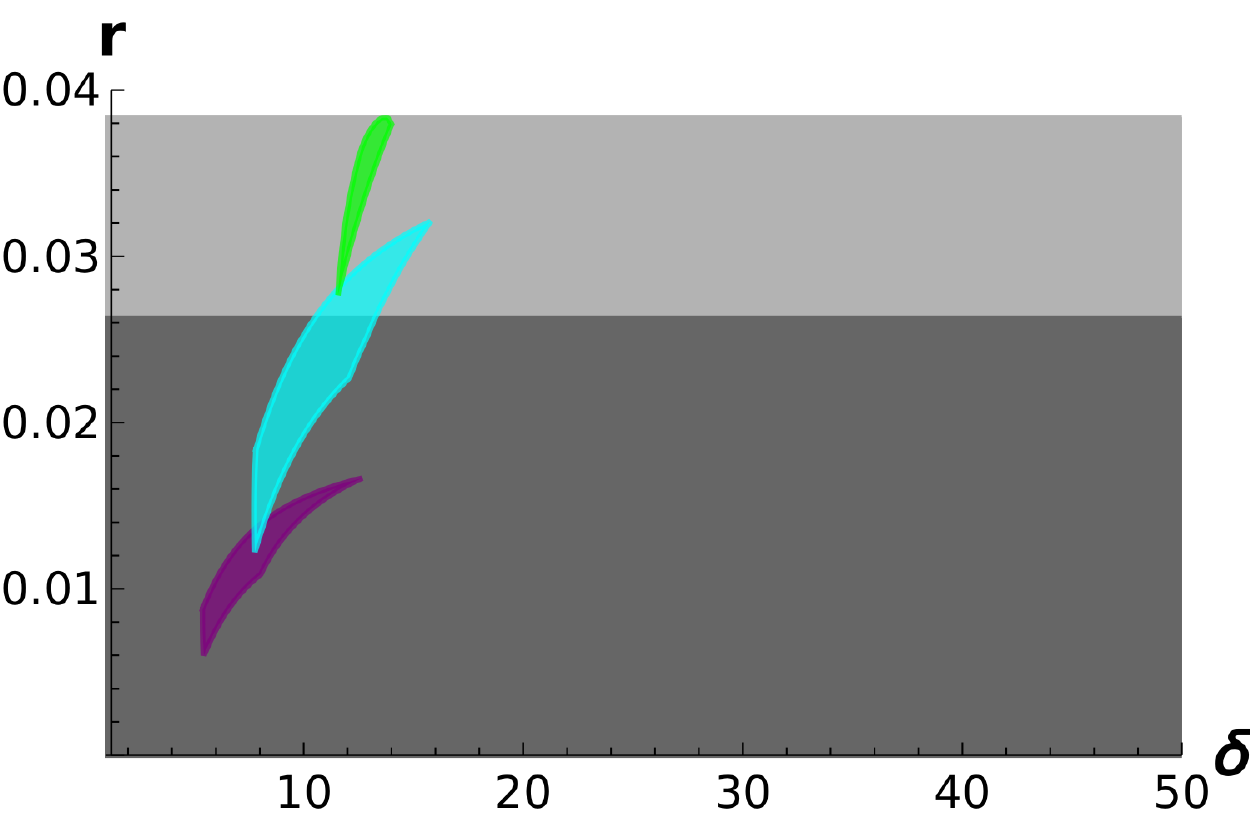}}%

    \subfloat[]{\includegraphics[width=0.5\textwidth]{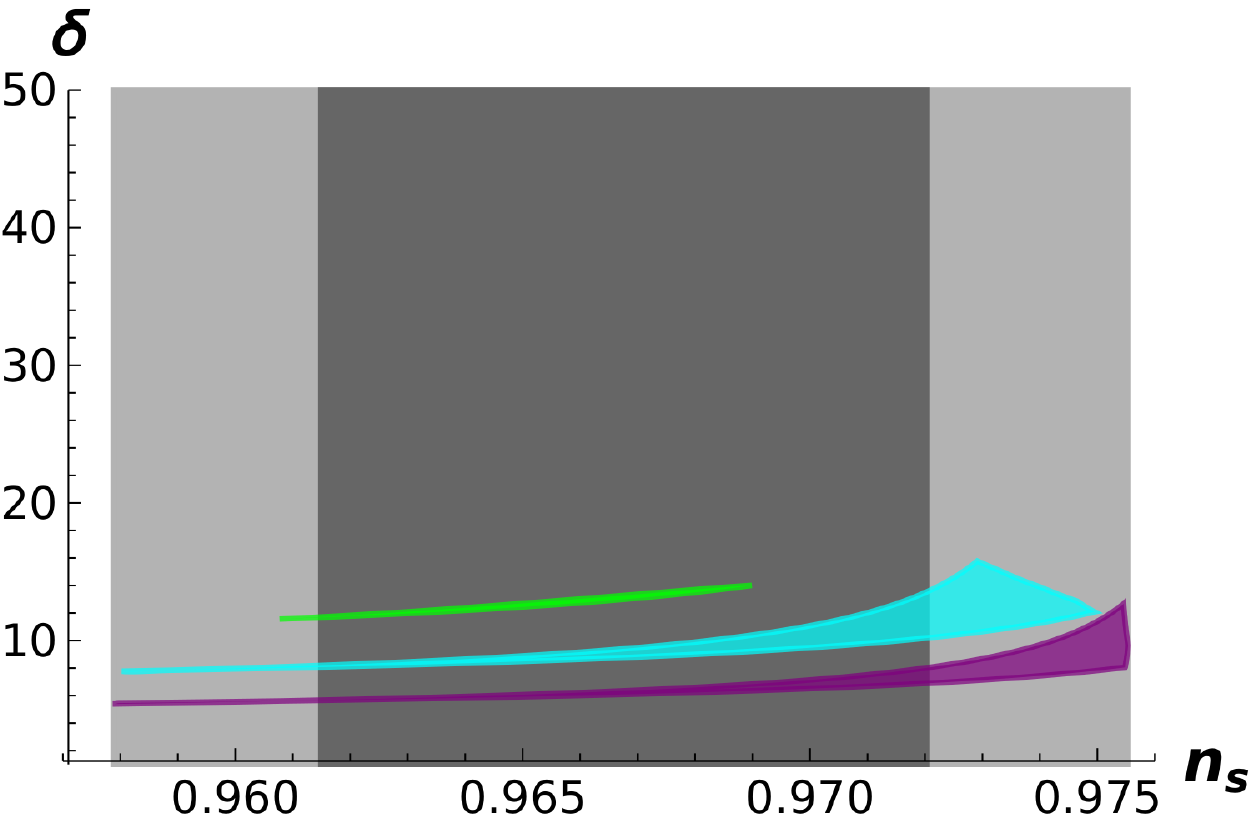}}%
    \subfloat[]{\includegraphics[width=0.5\textwidth]{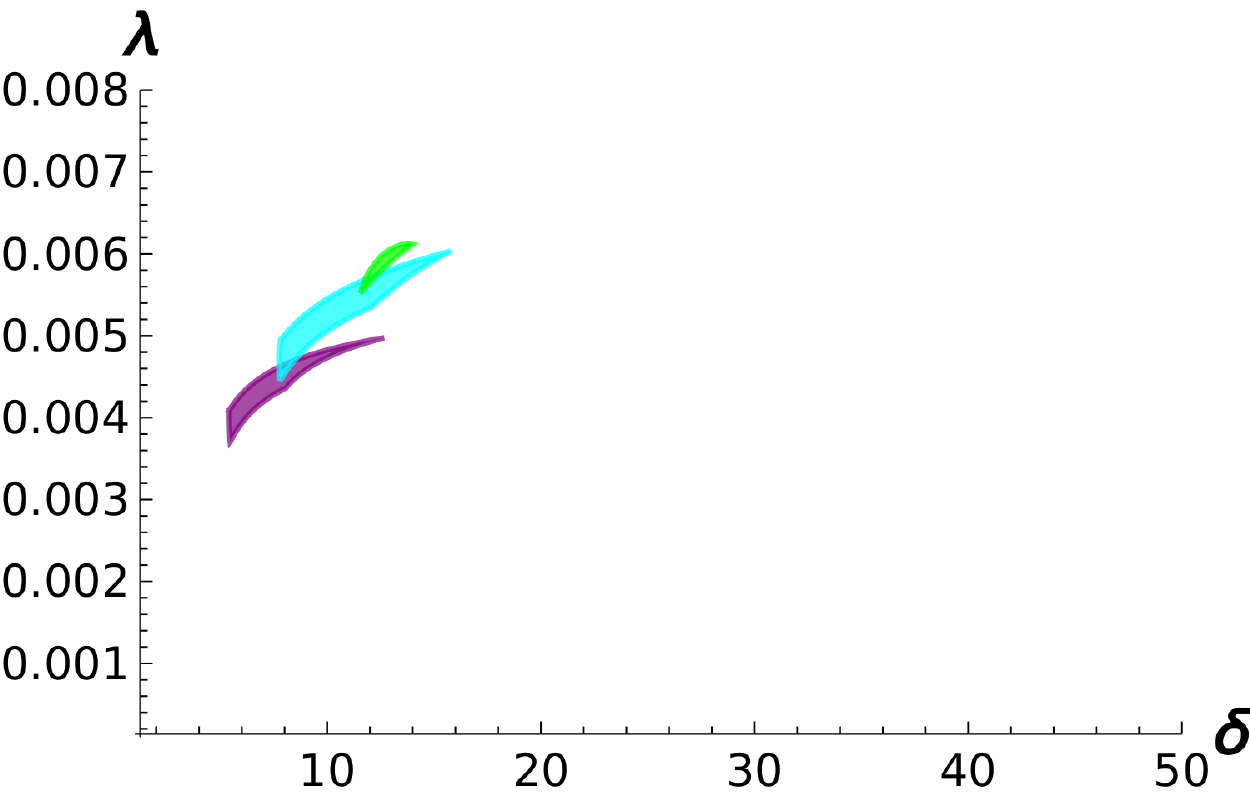}}%

    \caption{$r$ vs. $n_\text{s}$ (a),  $r$ vs. $\delta$ (b),  $\delta$ vs. $n_\text{s}$ (c),  $\lambda$ vs. $\delta$ (d). Filtered plots, where only the points inside the experimental constraints for $r$ and $n_\text{s}$ are kept for $m = \frac{3}{2}$ and $N_* \in [50, 60]$, except for the model with $n = 1$ (green) for which $N_* \in [56.6, 60]$. The models with $n = 2, 3$ and $4$ (blue, yellow and red in Fig. \ref{fig:m1.5}) are strongly disfavoured and therefore are missing. Green areas represent $n = 1$, cyan $n = \frac{1}{2}$ and purple $n = \frac{1}{4}$. With grey are the 1,2$\sigma$ allowed regions from the latest combination of Planck, BICEP/Keck and BAO data \cite{BICEP:2021xfz}.  } %
    \label{fig:m1.5_filt}%
\end{figure}


\subsubsection{Filtered results}

Fig. \ref{fig:m1.5_filt} shows the same plots as Fig. \ref{fig:m1.5}, but only for the points that are inside the experimental constraints for $r$ and $n_\text{s}$ \cite{BICEP:2021xfz}. Such constraints are explicitly visible in Fig. \ref{fig:m1.5_filt}(a), but the same filtering is also kept for the other plots. For instance, in Fig. \ref{fig:m1.5_filt}(b)  the purple and cyan areas seem to stop abruptly despite being well within the constraints for $r$ because they are no longer within the constraints for $n_\text{s}$ for those values of $r$. 

From Fig. \ref{fig:m1.5_filt}(a) it can be seen that the model with $n = 2$ (blue) is strongly disfavoured. The model with $n = 1$ (green) has some values of $\delta$ inside the 95\% confidence interval and none inside the 68\% interval. Moreover, in this configuration, there is a lower limit for the number of $e$-folds which is approximately $N_* \gtrsim 56.6$. 
The two other models ($n = \frac{1}{2}$ and $n = \frac{1}{4}$) both have several values of $\delta$ for which they are inside the 68\% confidence interval and they present no additional restriction on $N_*$.

Fig. \ref{fig:m1.5_filt}(b) shows the projection of the allowed areas for the models on the $r$ vs. $\delta$ plot. It can be seen that the areas within the experimental constraints are those that had a roughly linear relation between $\delta$ and $r$. Both the large values for $\delta$, where $r$ reached its asymptotic limit, as well as low values for $\delta$, where $r$ increased very rapidly, have been left out. While the slopes are similar, they are not exactly the same. Moreover, it can be noticed that the larger the value for $n$, the faster $r$ rises with an increase in $\delta$. Also, the smaller the value of $n$, the smaller values of $\delta$ are needed to obtain allowed solutions. For $n = 1$ (green), the valid range for $\delta$ is roughly between 12 and 14, for $n = \frac{1}{2}$ (cyan) the valid range for $\delta$ is roughly between 8 and 16 and for $n = \frac{1}{4}$ (purple) the valid range for $\delta$ is roughly between 5 and 13.

Fig. \ref{fig:m1.5_filt}(c) depicts the allowed regions on the $\delta$ vs. $n_\text{s}$ plot. This time the values of $\delta$ that are within the experimental constraints are such that for the smaller $\delta$ values the value of $n_\text{s}$ increases very quickly at first, but then at large values of $\delta$ the relation between $n_\text{s}$ and $\delta$ becomes nearly linear. An exception is the model with $n = 1$ (green), which is only valid in a small $\delta$ region, and the relationship is roughly linear. Moreover, for small values of $n_\text{s}$, there is a very small range of $\delta$ that is valid (i.e. the band is very narrow vertically at small $n_\text{s}$), but the larger the value of $n_\text{s}$, the larger the range of $\delta$, which can give that $n_\text{s}$ (i.e. the band is thick vertically at large $n_\text{s}$). Again, the smaller the value of $n$ the smaller values of $\delta$ are needed to provide valid solutions. The valid ranges for $\delta$ values are the same as in Fig. \ref{fig:m1.5_filt}(b) for each model.

In Fig. \ref{fig:m1.5_filt}(d) the valid areas for $\lambda$ vs. $\delta$ are plotted. For the model with $n = \frac{1}{4}$ (purple), for smaller $\delta$ values, $\lambda$ values increase at a very slow pace with the increase in $\delta$, and for larger $\delta$ the increase is even slower. For the model with $n = \frac{1}{2}$ (cyan), similar behaviour can be spotted, but the initial increase is slightly faster. The slope near the end of valid $\delta$ values is roughly the same, but the end itself is at larger values of $\delta$. The range of valid $\delta$ is rather small for the model with $n = 1$ (green) and around $\lambda \simeq 0.006$. Considering all $n$'s, the total allowed range for $\lambda$ is approximately between 0.004 and 0.006. Again, the valid $\delta$ regions are the same as in the previous plots in Fig. \ref{fig:m1.5_filt}(b,c).

\subsection{Model with $m = 2$}

\subsubsection{Unfiltered results}
\begin{figure}[t]
    \centering
     \subfloat[]{\includegraphics[width=0.5\textwidth]{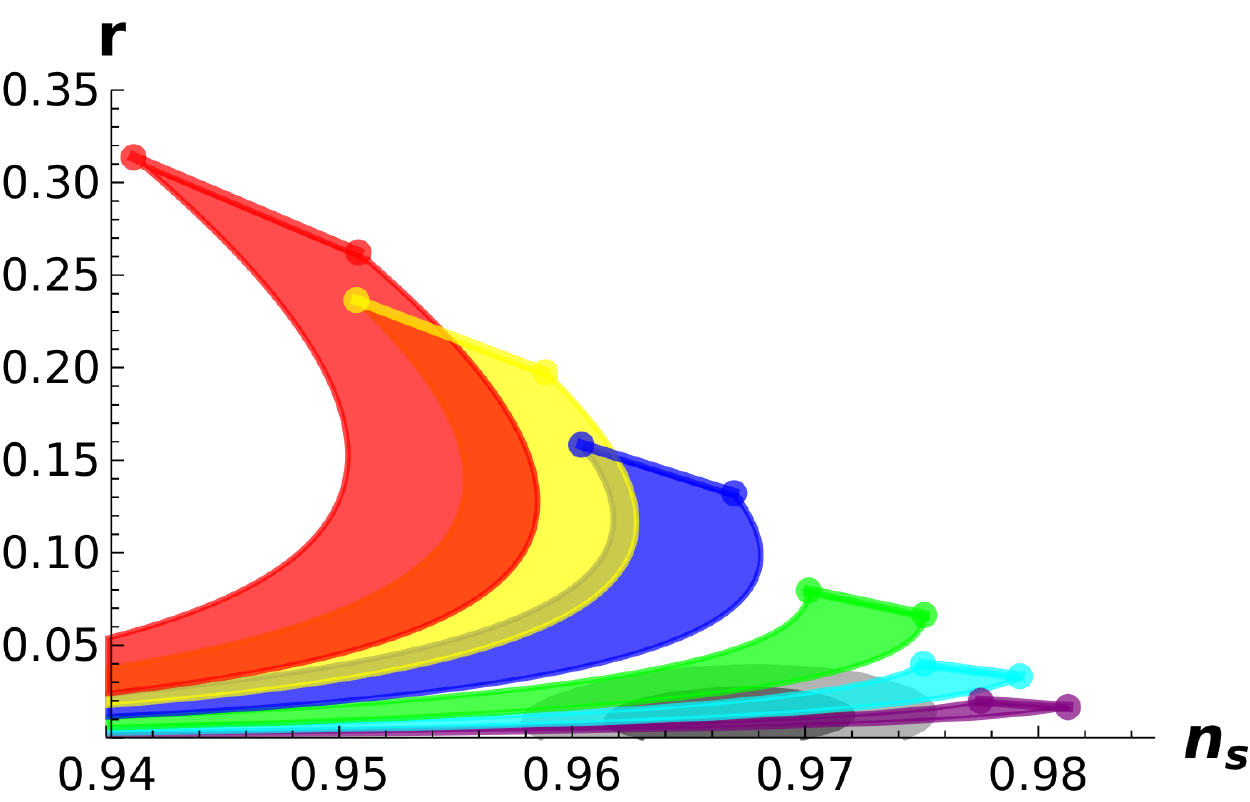}}%
    \subfloat[]{\includegraphics[width=0.5\textwidth]{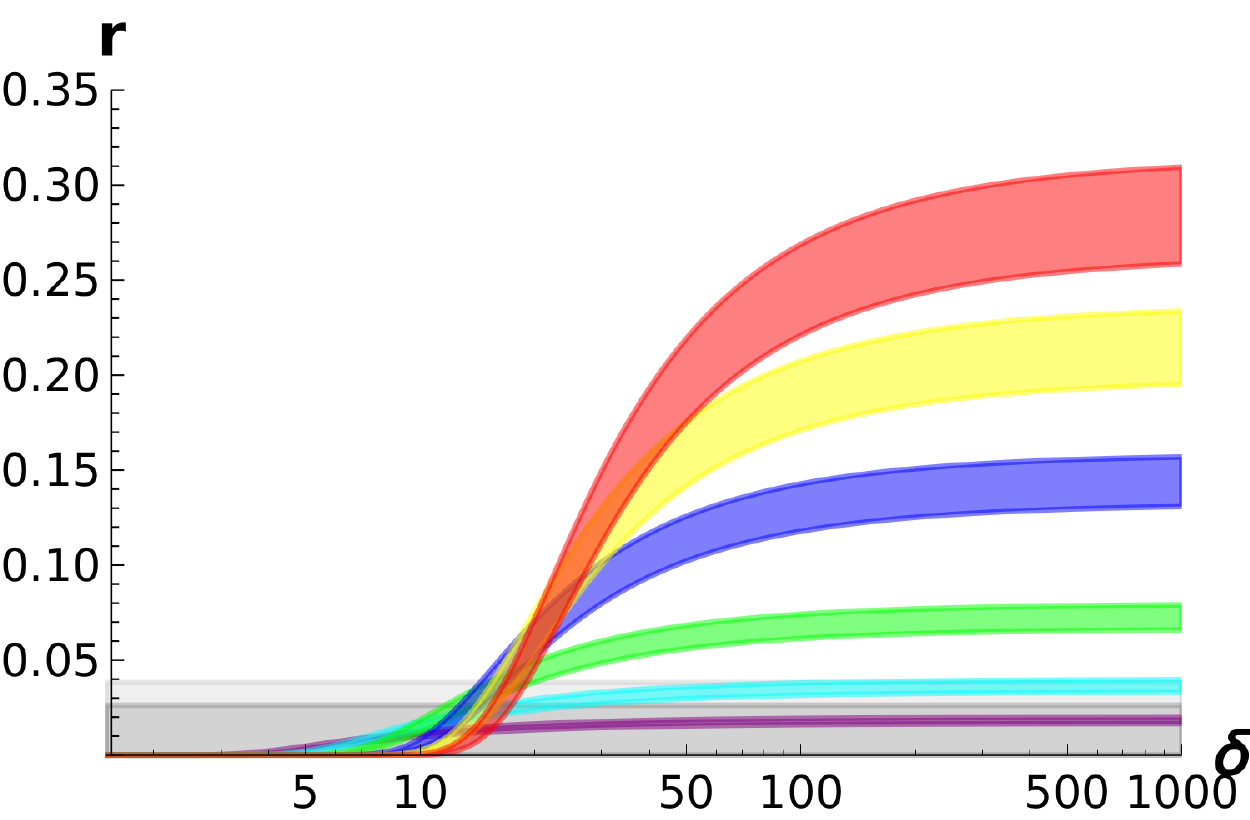}}%

    \subfloat[]{\includegraphics[width=0.5\textwidth]{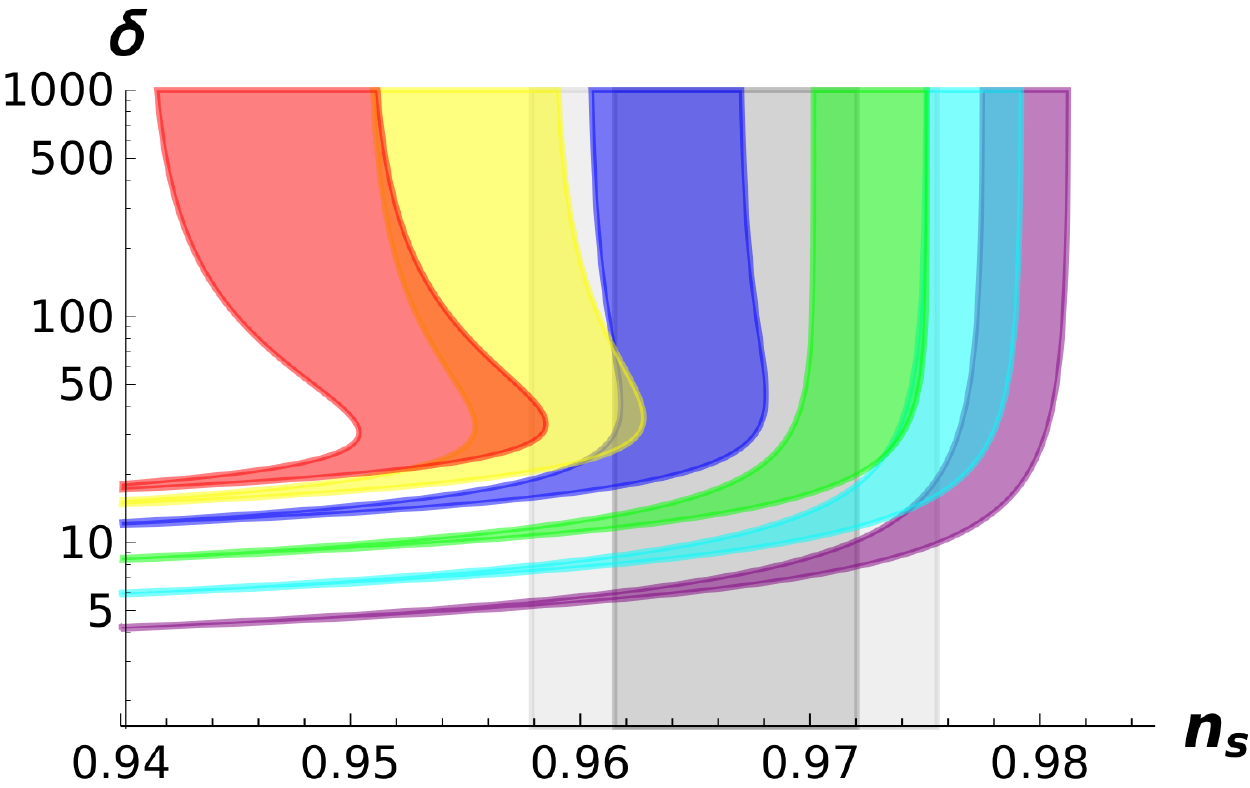}}%
    \subfloat[]{\includegraphics[width=0.5\textwidth]{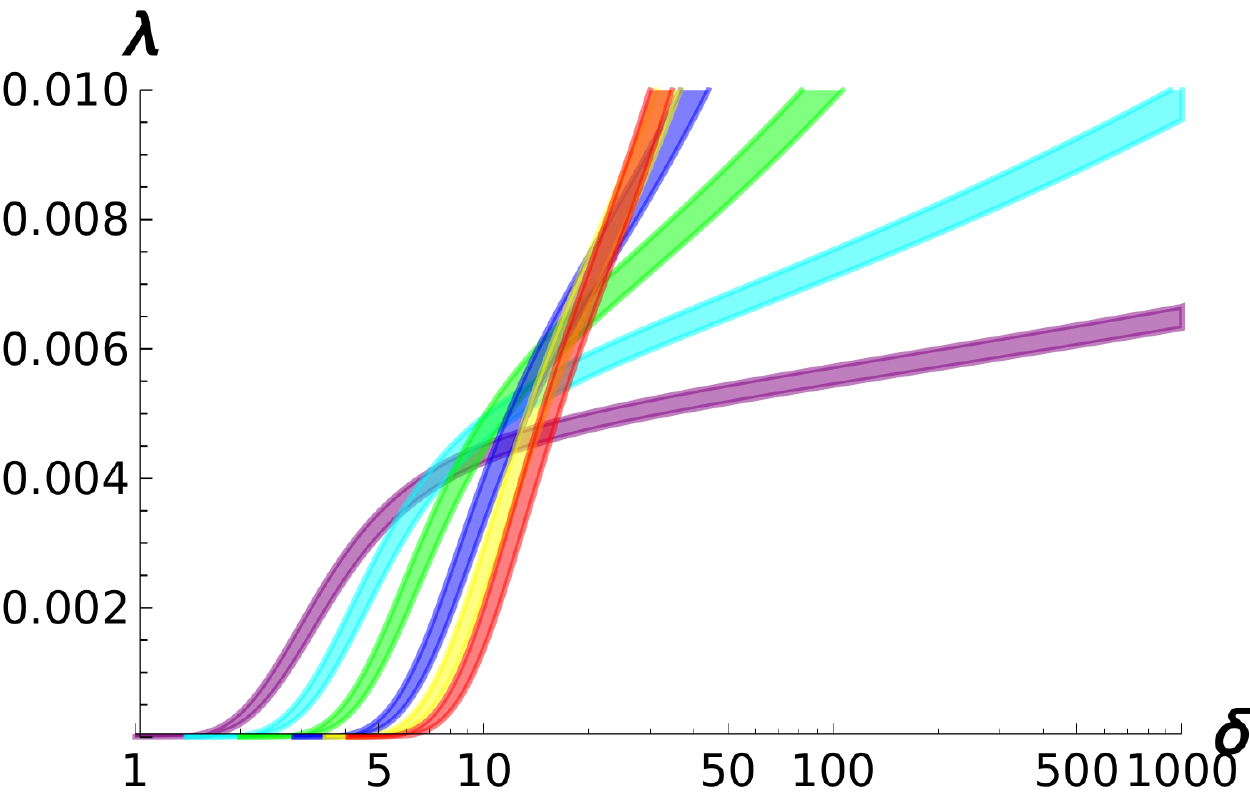}}%

    \caption{$r$ vs. $n_\text{s}$ (a),  $r$ vs. $\delta$ (b),  $\delta$ vs. $n_\text{s}$ (c),  $\lambda$ vs. $\delta$ (d). For these plots $m = 2$ and $N_* \in [50, 60]$. Red bands represent the model with $n = 4$, yellow $n = 3$, blue $n = 2$, green $n = 1$, cyan $n = \frac{1}{2}$ and purple $n = \frac{1}{4}$. In a continuous line, between bullets, with the same colour code, there are the corresponding monomial inflation limits.  With grey are the 1,2$\sigma$ allowed regions coming from the latest combination of Planck, BICEP/Keck and BAO data \cite{BICEP:2021xfz}. } %
    \label{fig:m2}%
\end{figure}

Fig. \ref{fig:m2} shows the same observables as Fig. \ref{fig:m1.5} but for $m=2$. The general shape for all plots with $m = 2$ is the same as for plots with $m = \frac{3}{2}$. The specific numerical values however are slightly different.

Fig. \ref{fig:m2}(a) shows $r$ vs. $n_\text{s}$ when $m = 2$. The shapes of the relations are roughly the same as for $m = \frac{3}{2}$, however, the bands are thicker and have only small gaps between them when $n$ is smaller or start to overlap significantly when $n$ is larger.

In Fig. \ref{fig:m2}(b) $r$ vs.  $\delta$ is plotted, when $m = 2$. While the general shape is the same as in Fig. \ref{fig:m1.5}(b), the bands are slightly thicker and the slopes at small $\delta$ are smaller.

In Fig. \ref{fig:m2}(c) $n_\text{s}$ vs.  $\delta$ is plotted, when $m = 2$. Just as with the previous plots, here the main difference with Fig. \ref{fig:m1.5} is the fact that the bands are slightly thicker. Also for $n \geq 2$ (blue, yellow and red) the decrease in $n_\text{s}$ at the large $\delta$ values is more visible.

In Fig. \ref{fig:m2}(d) $\lambda$ vs.  $\delta$ is plotted, when $m = 2$. Once again the general shape is similar to that in Fig. \ref{fig:m1.5}(d), however at smaller(larger) $\delta$ values $\lambda$ is slightly larger(smaller). In this case, when $\delta \lesssim  30$, $\lambda \lesssim  0.01$ (or even much smaller) for all the considered cases. 

\begin{figure}[t]
    \centering
     \subfloat[]{\includegraphics[width=0.5\textwidth]{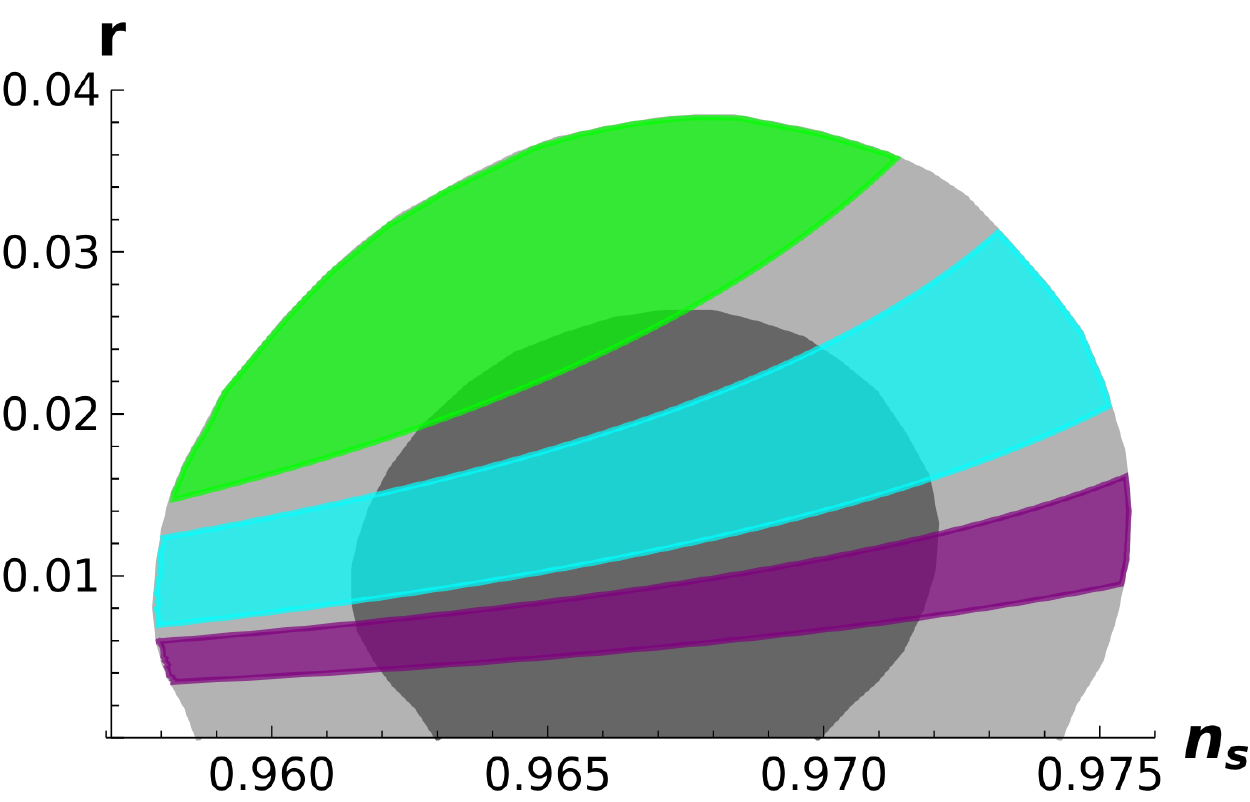}}%
    \subfloat[]{\includegraphics[width=0.5\textwidth]{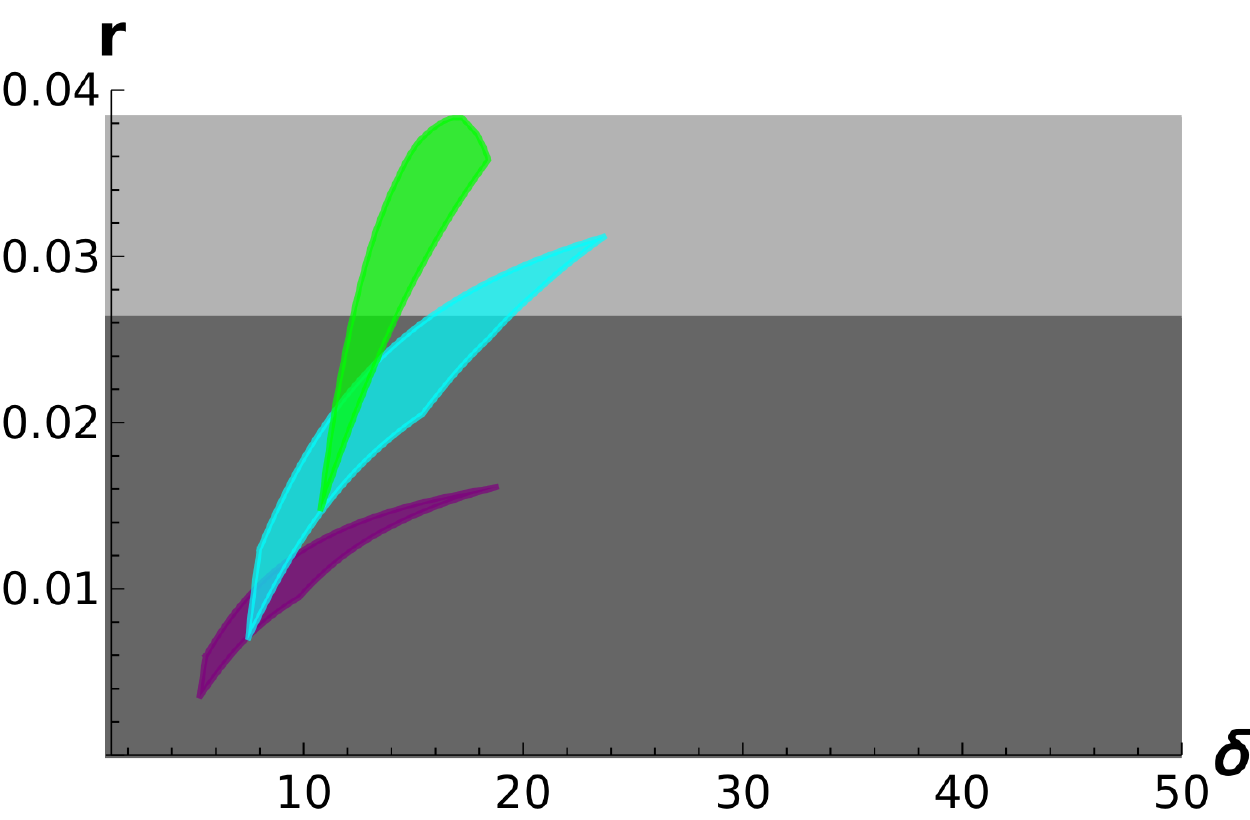}}%

    \subfloat[]{\includegraphics[width=0.5\textwidth]{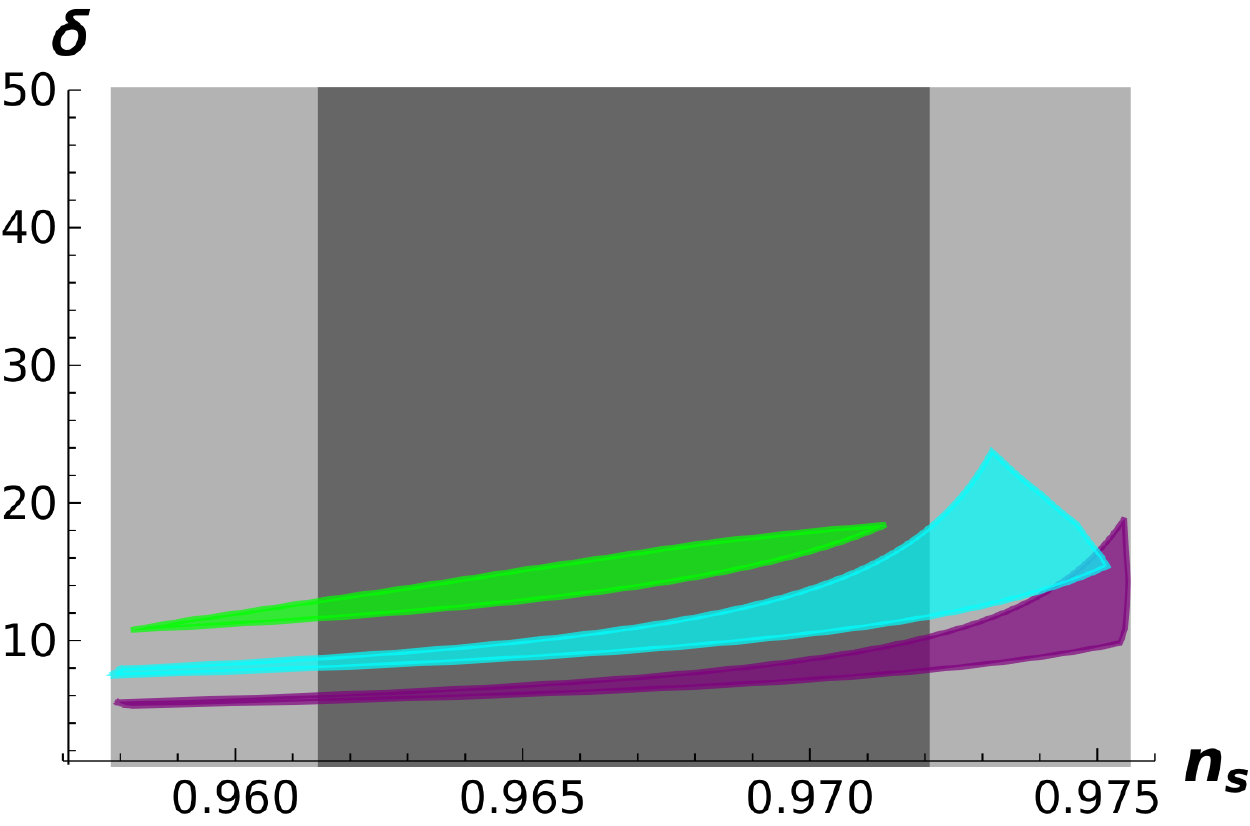}}%
    \subfloat[]{\includegraphics[width=0.5\textwidth]{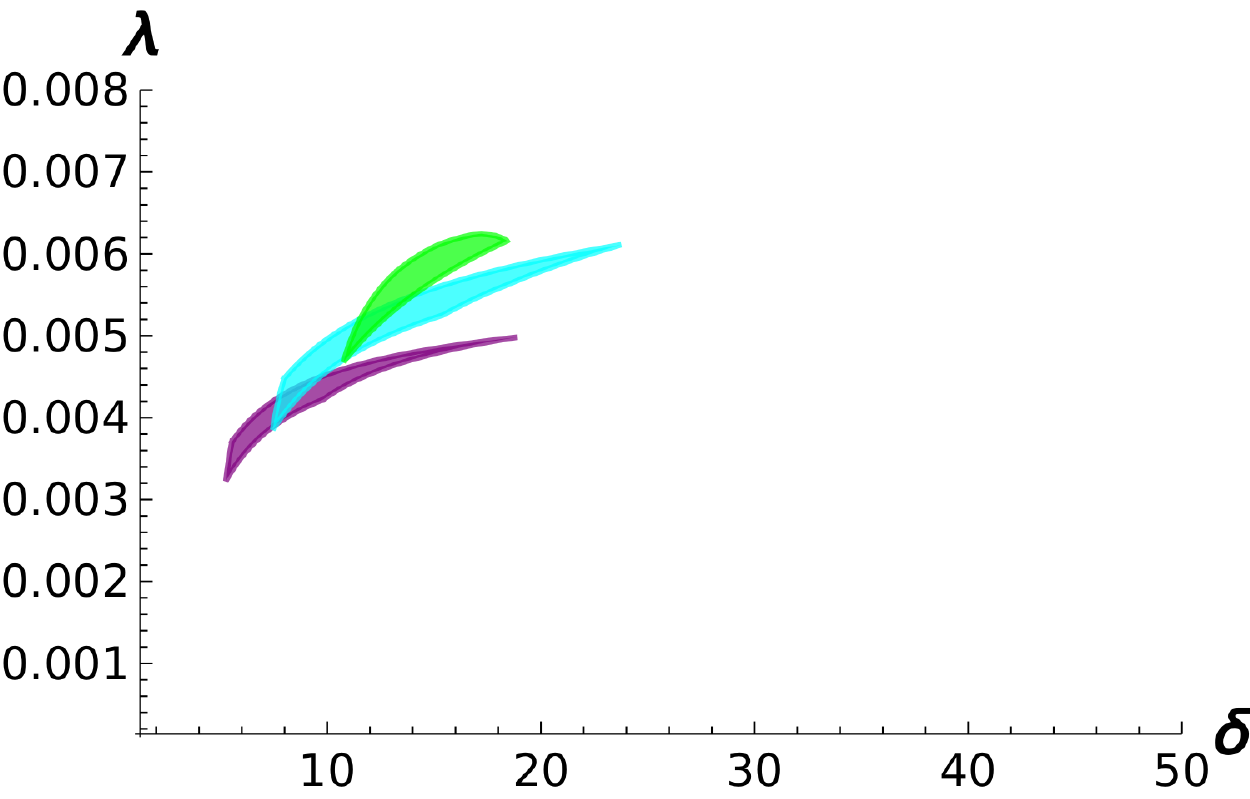}}%

     \caption{$r$ vs. $n_\text{s}$ (a),  $r$ vs. $\delta$ (b),  $\delta$ vs. $n_\text{s}$ (c),  $\lambda$ vs. $\delta$ (d). Filtered plots, where only the points inside the experimental constraints for $r$ and $n_\text{s}$ are kept for $m = 2$ and $N_* \in [50, 60]$, except for the model with $n = 1$ (green) for which $N_* \in [52, 60]$. The models with $n = 2, 3$ and $4$ (blue, yellow and red in Fig. \ref{fig:m2}) are strongly disfavoured and therefore are missing. The green areas represent the model with $n = 1$, cyan $n = \frac{1}{2}$ and purple $n = \frac{1}{4}$. With grey are the 1,2$\sigma$ allowed regions from the latest combination of Planck, BICEP/Keck and BAO data \cite{BICEP:2021xfz}. }
    \label{fig:m2_filt}%
\end{figure}

\subsubsection{Filtered results}
The plots in Fig. \ref{fig:m2_filt} follow the same logic as the ones in Fig. \ref{fig:m1.5_filt} (i.e. they are the same as Fig. \ref{fig:m2}, but only points in 2$\sigma$ threshold for $r$ and $n_\text{s}$ constraints are kept and they have been zoomed in). The general shapes are also the same as in Fig. \ref{fig:m1.5_filt}, but the range of valid $\delta$ is larger.

From Fig. \ref{fig:m2_filt}(a) it can be seen that the model with $n = 2$ (blue) is immediately strongly disfavoured as expected since this configuration is nothing but the symmetry breaking model \cite{OLIVE1990307}. The model with $n = 1$ (green) has some values of $\delta$ for which it is inside the 95\% confidence interval and a few for which it is inside the 68\% confidence interval. Both, the model with $n = \frac{1}{2}$ (cyan) and $n = \frac{1}{4}$ (purple), have some values of $\delta$ for which they are inside the 68\% confidence interval. And as mentioned before, for $n = \frac{1}{4}$ and $n = \frac{1}{2}$, there are valid solutions for all $N_* \in [50, 60]$, but for $n = 1$, there are valid solutions for only $N_* \in [52, 60]$. 

In Fig. \ref{fig:m2_filt}(b) the general shape of the valid regions is roughly the same as in Fig. \ref{fig:m1.5_filt}(b) and just as before, the slope for $\delta$ vs. $r$ is smaller for smaller values of $n$ and vice versa. However, the ranges of valid $\delta$ are significantly larger for $m = 2$, than for $m = \frac{3}{2}$. The valid ranges of $\delta$ are roughly from 5 to 19 for $n = \frac{1}{4}$ (purple), from 7 to 24 for $n = \frac{1}{2}$ (cyan) and from 11 to 18 for $n = 1$ (green).

Fig. \ref{fig:m2_filt}(c) depicts $\delta$ vs. $n_\text{s}$ in the valid region, with parameter $m$ being 2. The bands are thicker and the increase in $n_\text{s}$ at the small $\delta$ values is not as large as in Fig. \ref{fig:m1.5_filt}(c). However, the general pattern, of having a smaller increase in $n_\text{s}$ per increase in $\delta$ as $\delta$ gets larger, still holds. For $n = \frac{1}{2}$ (cyan), the increase in $n_\text{s}$ at larger values of $\delta$ almost stops (i.e. the upper edge of the cyan plot becomes nearly vertical).

Fig. \ref{fig:m2_filt}(d) depicts $\lambda$ vs. $\delta$ in the valid region, with parameter $m$ being 2. The shapes for different models are roughly the same as in Fig. \ref{fig:m1.5_filt}(d) and the extension of all allowed areas is now larger. Considering all $n$'s, the highest valid values for $\lambda$ is still around 0.006 while the smallest is now around 0.003.


\begin{figure}[t]
    \centering
     \subfloat[]{\includegraphics[width=0.5\textwidth]{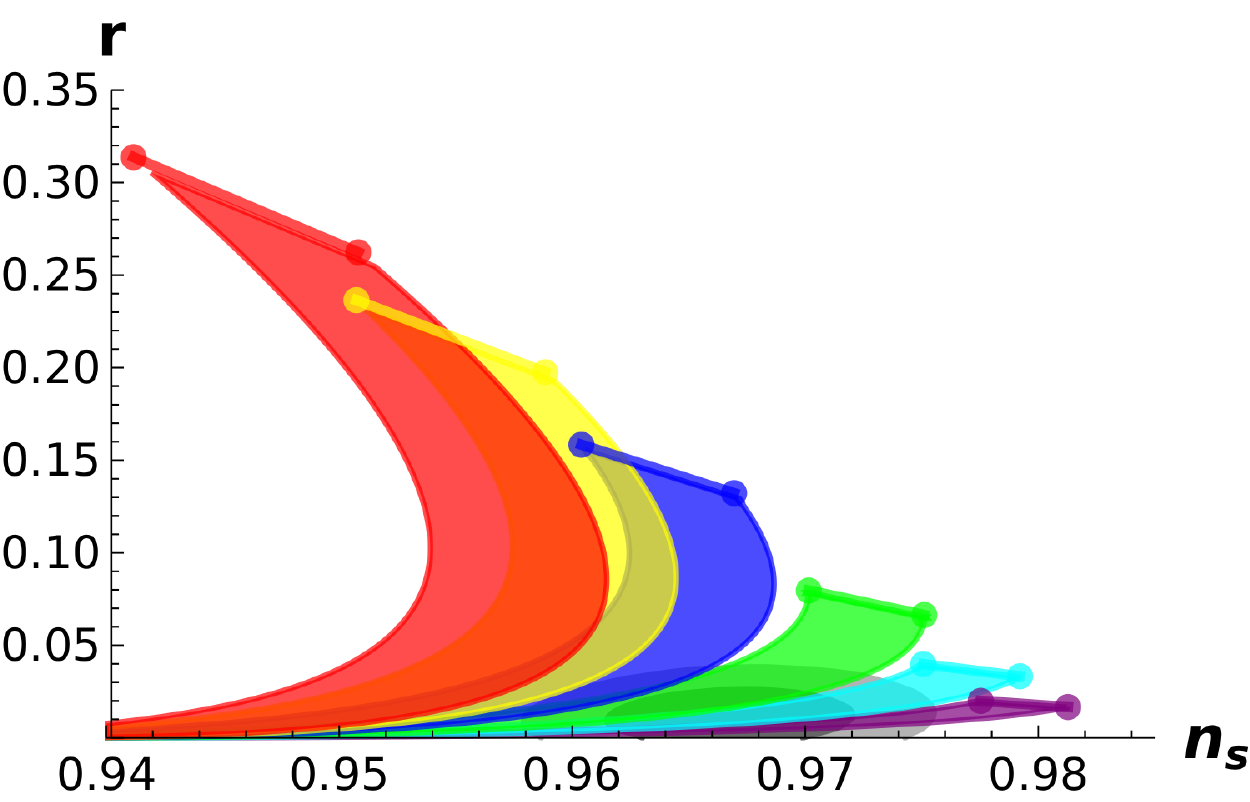}}%
    \subfloat[]{\includegraphics[width=0.5\textwidth]{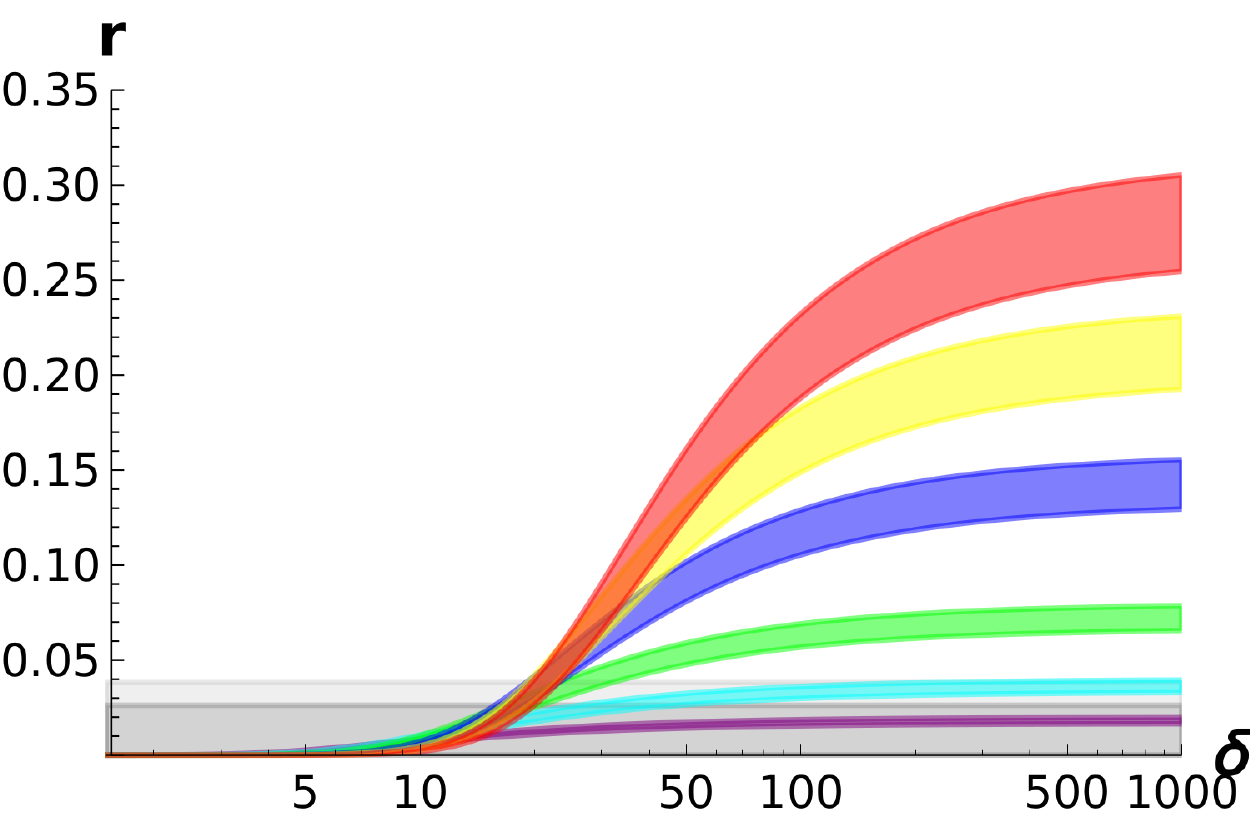}}%

    \subfloat[]{\includegraphics[width=0.5\textwidth]{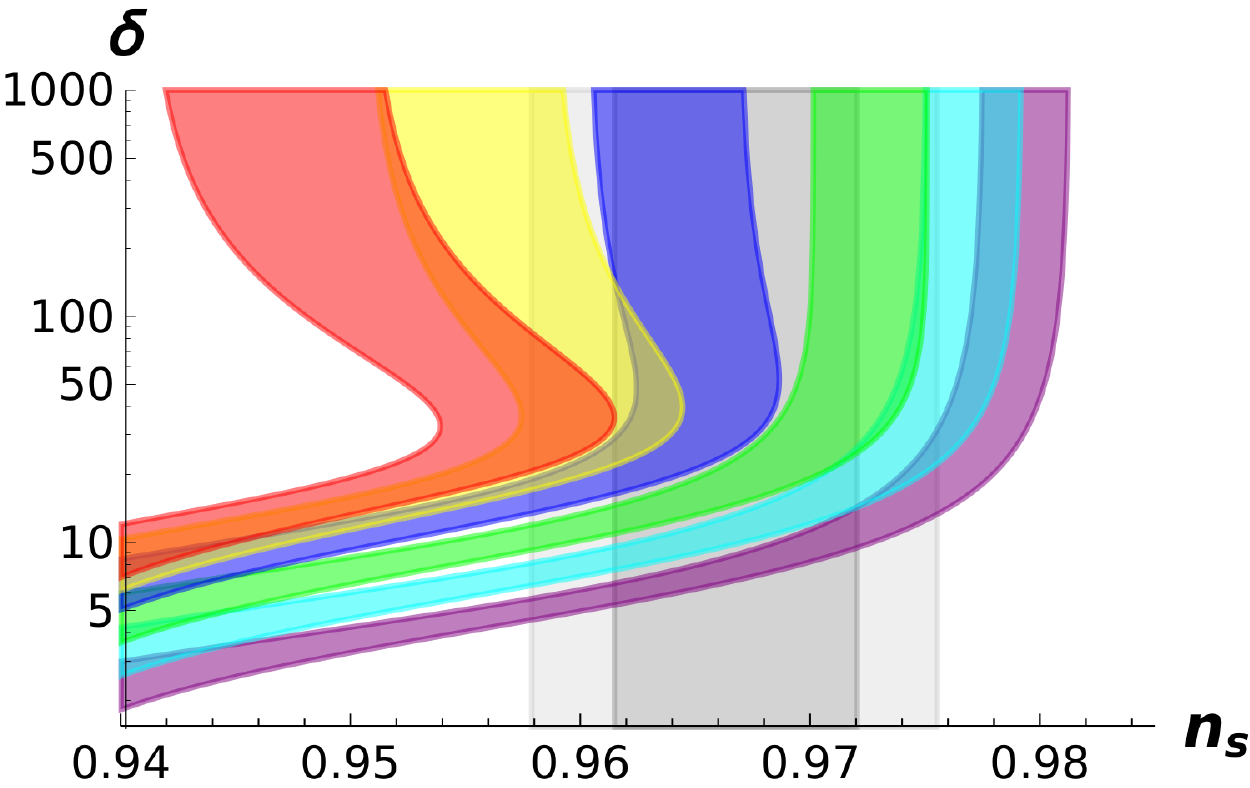}}%
    \subfloat[]{\includegraphics[width=0.5\textwidth]{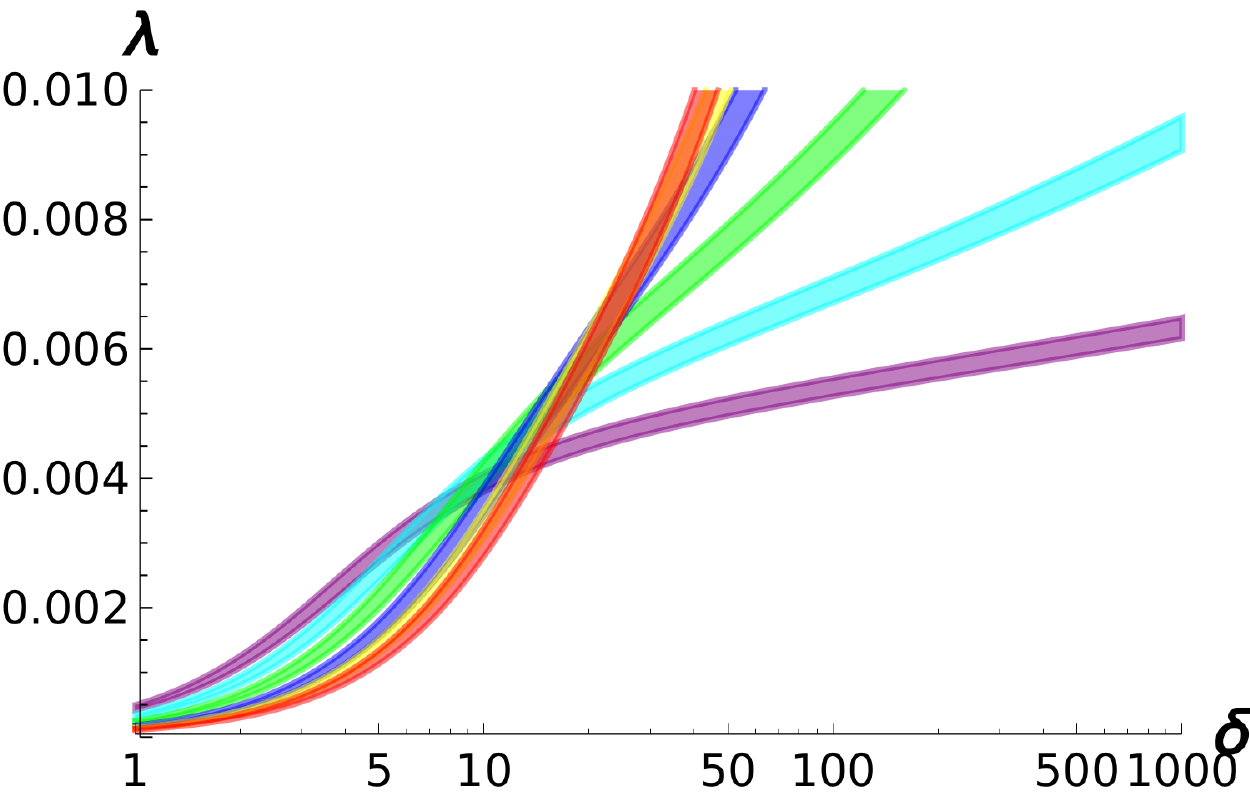}}%

    \caption{$r$ vs. $n_\text{s}$ (a),  $r$ vs. $\delta$ (b),  $\delta$ vs. $n_\text{s}$ (c),  $\lambda$ vs. $\delta$ (d). For these plots $m = 3$ and $N_* \in [50, 60]$. Red bands represent the model with $n = 4$, yellow $n = 3$, blue $n = 2$, green $n = 1$, cyan $n = \frac{1}{2}$ and purple $n = \frac{1}{4}$. In a continuous line, between bullets, with the same colour code, there are the corresponding monomial inflation limits.  With grey are the 1,2$\sigma$ allowed regions coming from the latest combination of Planck, BICEP/Keck and BAO data \cite{BICEP:2021xfz}. } %
    \label{fig:m3}%
\end{figure}

\begin{figure}[t]
    \centering
     \subfloat[]{\includegraphics[width=0.5\textwidth]{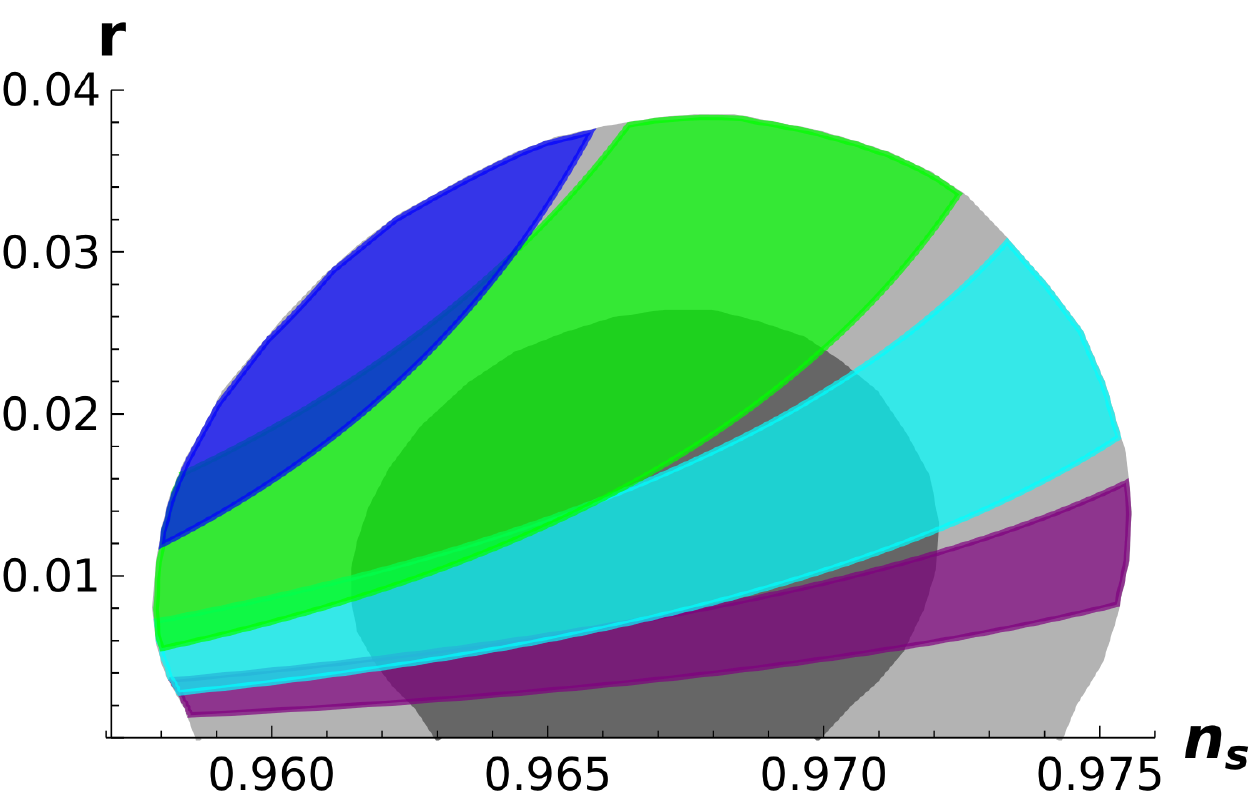}}%
    \subfloat[]{\includegraphics[width=0.5\textwidth]{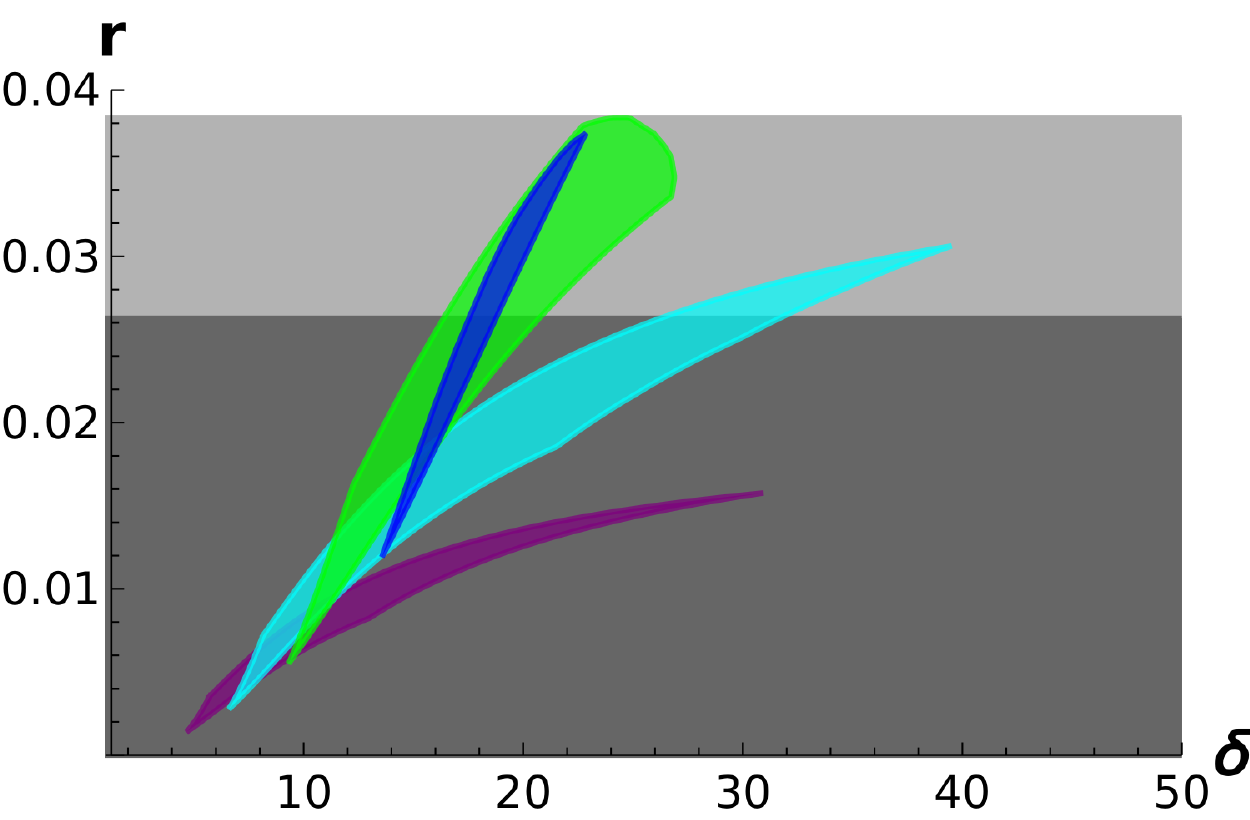}}%

    \subfloat[]{\includegraphics[width=0.5\textwidth]{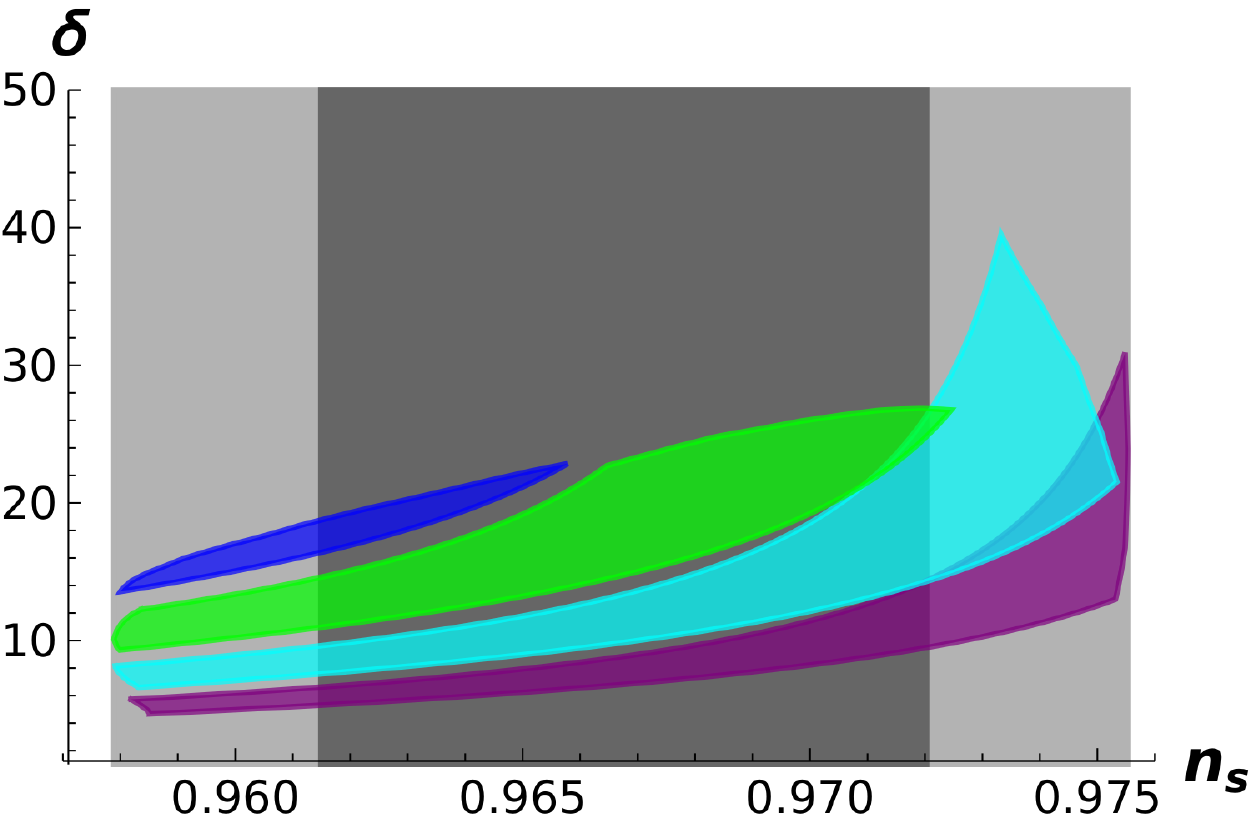}}%
    \subfloat[]{\includegraphics[width=0.5\textwidth]{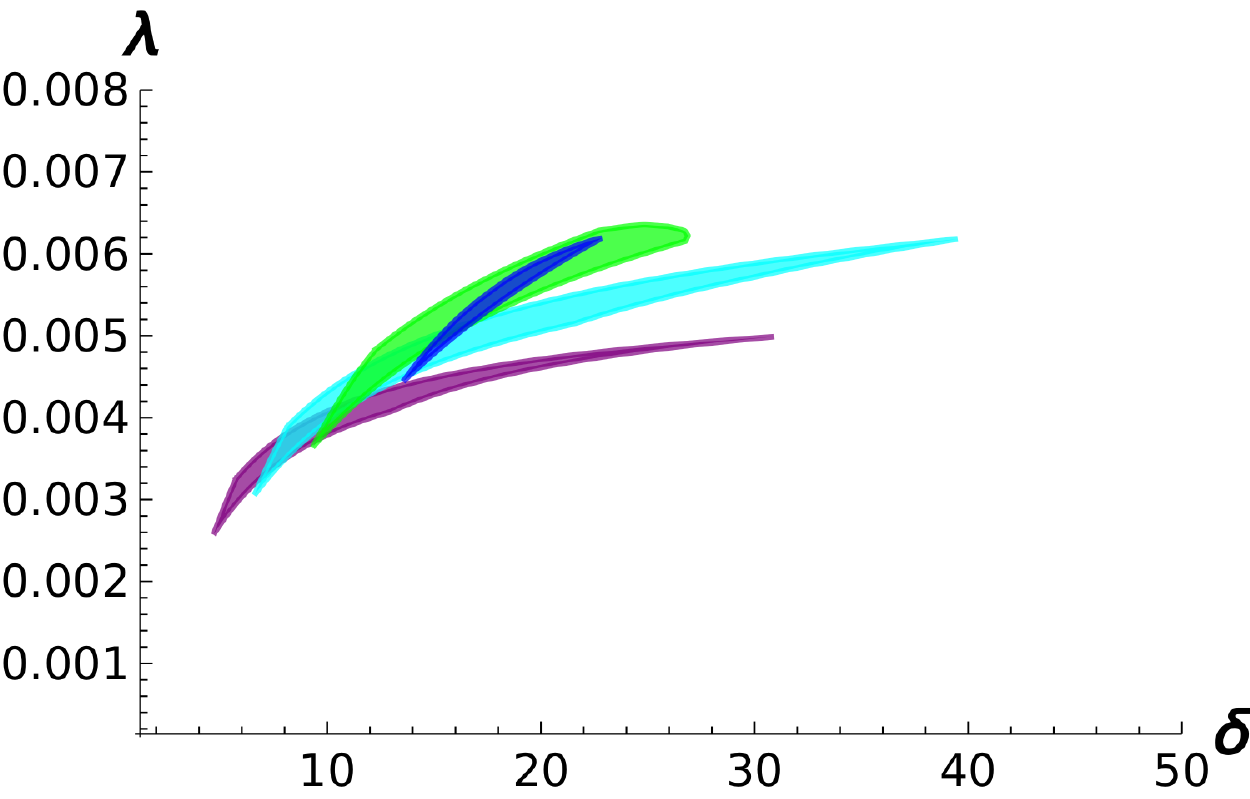}}%

     \caption{$r$ vs. $n_\text{s}$ (a),  $r$ vs. $\delta$ (b),  $\delta$ vs. $n_\text{s}$ (c),  $\lambda$ vs. $\delta$ (d). Filtered plots, where only the points inside the experimental constraints for $r$ and $n_\text{s}$ are kept for $m = 3$ and $N_* \in [50, 60]$. The blue areas represent $n = 2$, green areas represent $n = 1$, cyan $n = \frac{1}{2}$ and purple $n = \frac{1}{4}$. With grey are the 1,2$\sigma$ allowed regions coming from the latest combination of Planck, BICEP/Keck and BAO data \cite{BICEP:2021xfz}. }
    \label{fig:m3_filt}%
\end{figure}

\subsection{Model with $m = 3$}
Fig. \ref{fig:m3} shows the same observables as Fig. \ref{fig:m1.5} and Fig. \ref{fig:m2} but for $m=3$. The general shape for all plots with $m = 3$ is similar to the previous unfiltered plots, however, there are some differences in the exact behaviour and also the numerical values are different.
\subsubsection{Unfiltered results}

Fig. \ref{fig:m3}(a) shows $r$ vs. $n_\text{s}$ when $m = 3$. The shapes of the relations are roughly the same as for $m = \frac{3}{2}$ and $m = 2$. The bands are, however, thicker (except for the red ones with $n = 4$) and for smaller $n$ values there are even smaller gaps between them compared to the plots with smaller $m$ values. Also, for bigger $n$ values the overlap is even more prominent compared to the overlap with bigger $n$ values in $m = 2$.

In Fig. \ref{fig:m3}(b) $r$ vs.  $\delta$ is plotted, when $m = 3$. While the general shape is the same as in Fig. \ref{fig:m1.5}(b) and Fig. \ref{fig:m2}(b), the bands are slightly thicker. The slopes at small $\delta$ are even smaller than for $m = 2$ with $r$ values increasing at a slower pace with the increase in $\delta$ values.

In Fig. \ref{fig:m3}(c) $n_\text{s}$ vs.  $\delta$ is plotted, when $m = 3$. The main difference compared to smaller $m$ values is that when for $m = \frac{3}{2}$ and $m = 2$ the graphs are almost horizontal for smaller $\delta$, then as for $m = 3$ they are gradually increasing with the increase in $n_\text{s}$. Also, the bands are generally thicker compared to plots with smaller $m$ values. Again, for $n \geq 2$ (blue, yellow and red) the decrease in $n_\text{s}$ at larger $\delta$ values is more visible.

Fig. \ref{fig:m3}(d) depicts $\lambda$ vs.  $\delta$, when $m = 3$. Again the general shape is similar to the previously studied case, and again at smaller(larger) $\delta$ values $\lambda$ is slightly larger(smaller). In this case, when $\delta \lesssim  40$, $\lambda \lesssim  0.01$ (or even much smaller) for all the considered cases. 

\subsubsection{Filtered results}
The plots in Fig. \ref{fig:m3_filt} follow the same logic as the ones in Fig. \ref{fig:m1.5_filt} and Fig. \ref{fig:m2_filt} (i.e. they are the same as Fig. \ref{fig:m3}, but only points in 2$\sigma$ threshold for $r$ and $n_\text{s}$ constraints are kept and they have been zoomed in). The general shapes are the same as in the previous filtered plots, but the valid areas are larger and this time the model with $n = 2$ (blue) is also valid.

From Fig. \ref{fig:m3_filt}(a) it can be seen that the models with $n = 3$ (yellow) and $n = 4$ (red) are immediately strongly disfavoured. The model with $n = 2$ (blue) is no longer completely disfavored with some values being inside the 95\% confidence interval.  The model with $n = 1$ (green) has many values inside the 95\% confidence interval and significantly more values of $\delta$ for which it is inside the 68\% confidence interval compared to  $m = \frac{3}{2}$ and $m = 2$. As for the models with $n = \frac{1}{2}$ (cyan) and $n = \frac{1}{4}$ (purple), they have slightly more values of $\delta$ for which they are inside the 68\% confidence interval compared to the previous $m $ values. 

In Fig. \ref{fig:m3_filt}(b) the general shape of the valid regions is roughly the same as for the filtered plots of $m = \frac{3}{2}$ and $m = 2$. One of the main differences is that for this $m$ value for combinations with $n = 1$ (green), $n=\frac{1}{2}$ (cyan) and $n=\frac{1}{4}$ (purple) the valid range of $\delta$ is larger, with bigger values valid as well. Secondly, for this $m$ there are also some valid points for $n = 2$ (blue).

Fig. \ref{fig:m3_filt}(c) depicts $\delta$ vs. $n_{\text{s}}$ in the valid region, with the parameter $m$ being 3. The bands are even thicker, with significantly more points inside the confidence intervals for all the valid $n$ values. Again, for this $m$, the model with $n = 2$ (blue) is also valid. The increase in $n_{\text{s}}$ at the small $\delta$ values is smaller compared to Fig. \ref{fig:m1.5_filt}(c) and \ref{fig:m2_filt}(c). With the increase in $\delta$ the increase in $n_{\text{s}}$ becomes slower. 

Fig. \ref{fig:m3_filt}(d) depicts $\lambda$ vs. $\delta$ in the valid region, with parameter $m$ being 3. Again, the shapes for different models are roughly the same as in Fig. \ref{fig:m1.5_filt}(d) and Fig. \ref{fig:m2_filt}(d), the extension of all allowed areas is even larger and also the case $n=2$ is allowed now. The highest valid values for $\lambda$ are slightly smaller than the corresponding ones in Fig. \ref{fig:m1.5_filt}(d) and Fig. \ref{fig:m2_filt}(d). Considering all $n$'s, the highest valid values for $\lambda$ is still around 0.006 while the smallest is now around 0.0025.

\subsection{Model with $m = 4$}

\subsubsection{Unfiltered results}
\begin{figure}[t]
    \centering
     \subfloat[]{\includegraphics[width=0.5\textwidth]{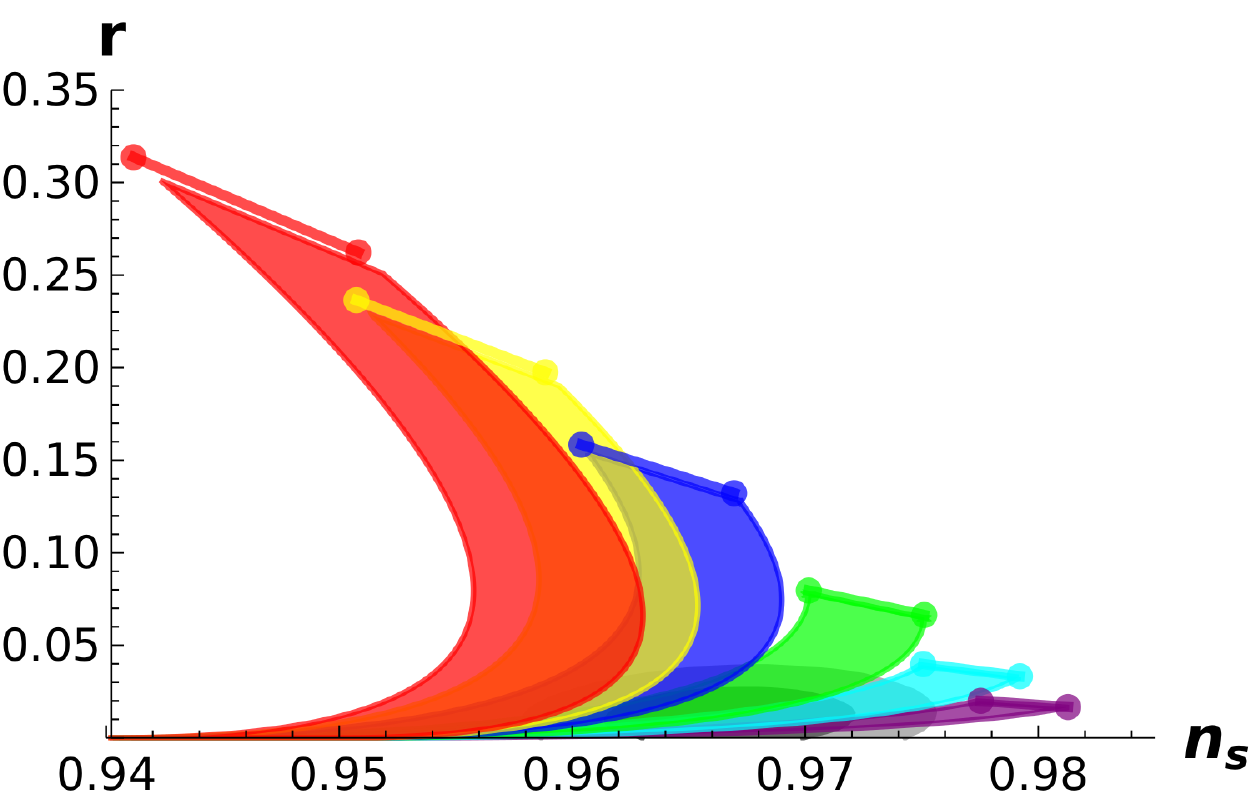}}%
    \subfloat[]{\includegraphics[width=0.5\textwidth]{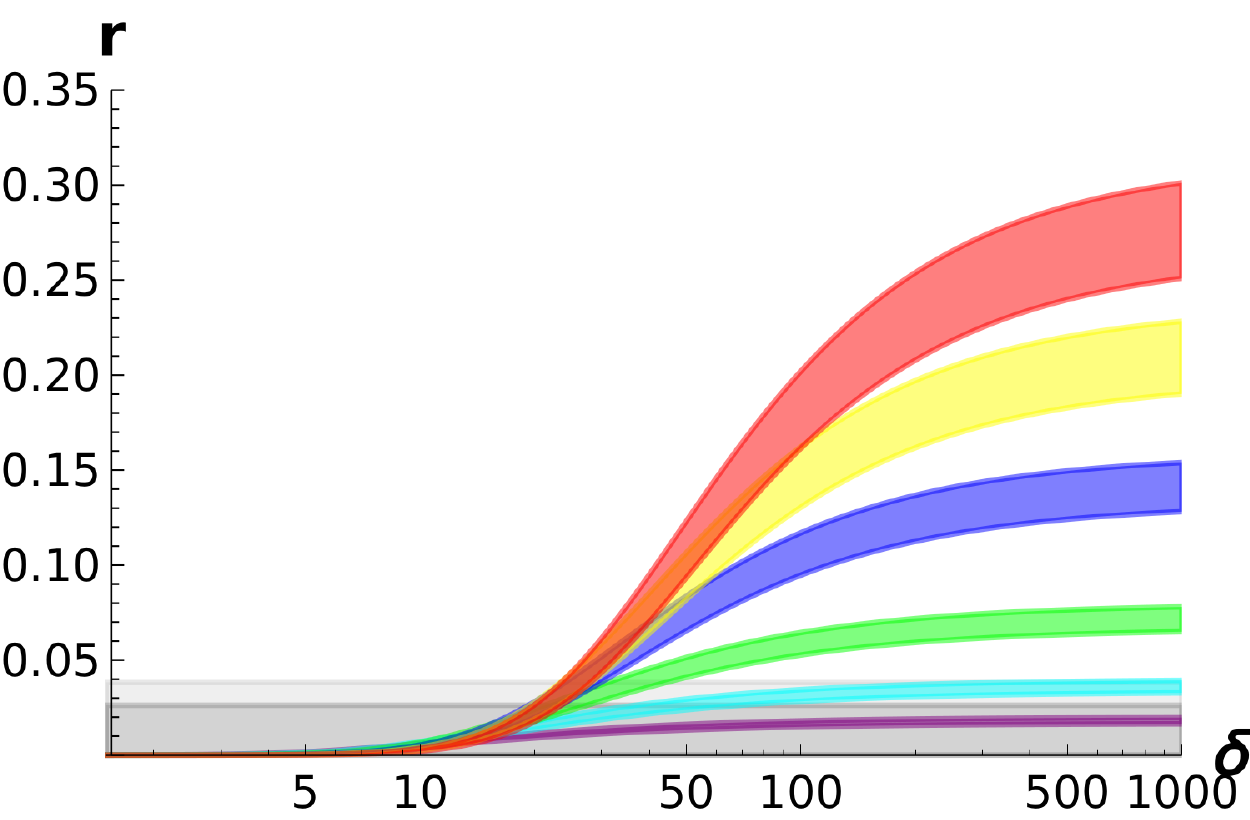}}%

    \subfloat[]{\includegraphics[width=0.5\textwidth]{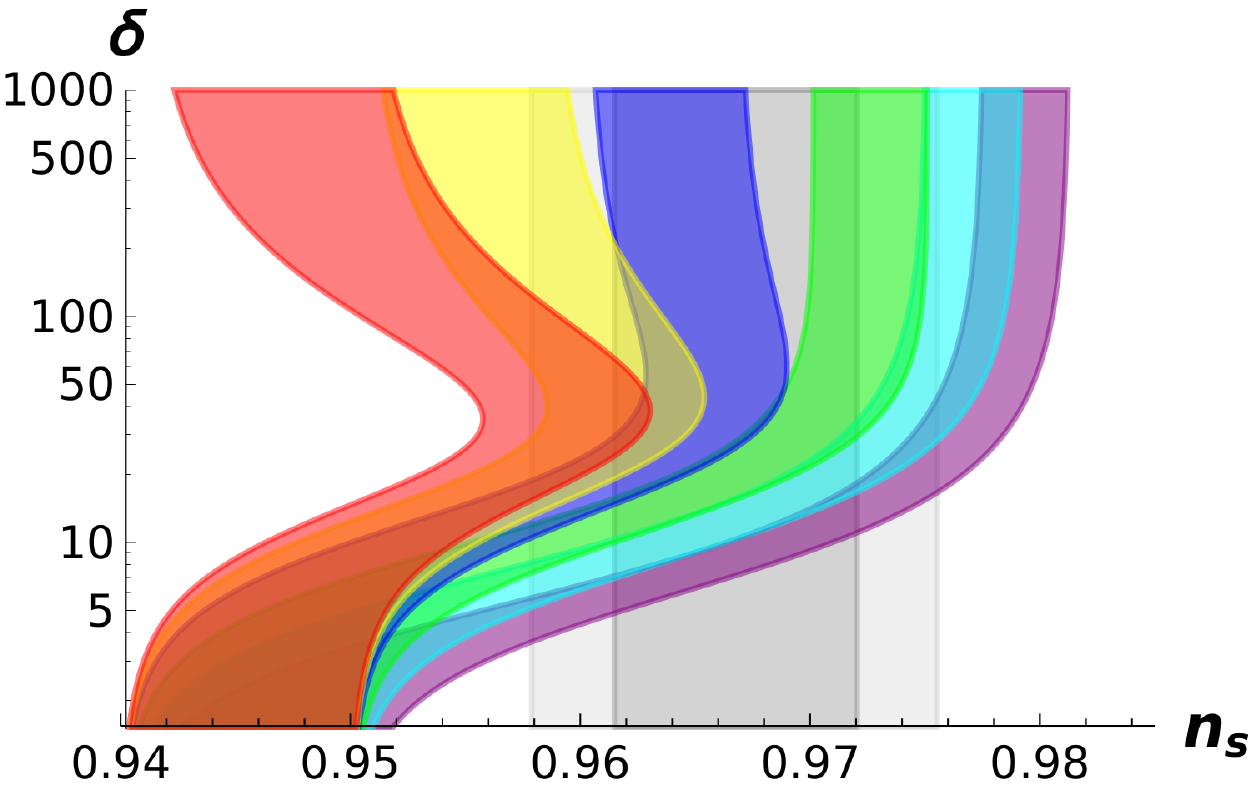}}%
    \subfloat[]{\includegraphics[width=0.5\textwidth]{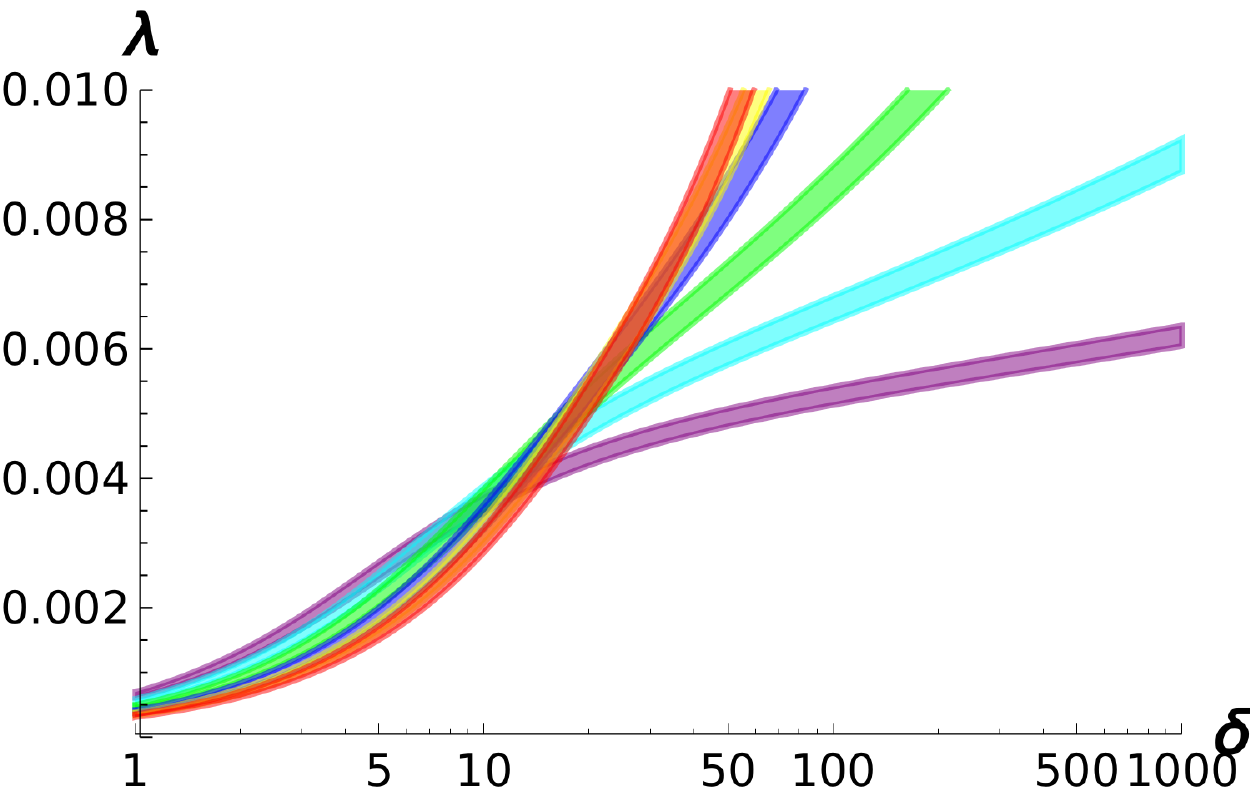}}%

    \caption{$r$ vs. $n_\text{s}$ (a),  $r$ vs. $\delta$ (b),  $\delta$ vs. $n_\text{s}$ (c),  $\lambda$ vs. $\delta$ (d). For these plots $m = 4$ and $N_* \in [50, 60]$. Red bands represent the model with $n = 4$, yellow $n = 3$, blue $n = 2$, green $n = 1$, cyan $n = \frac{1}{2}$ and purple $n = \frac{1}{4}$. There are the corresponding monomial inflation limits in a continuous line between bullets with the same colour code.  With grey are the 1,2$\sigma$ allowed regions from the latest combination of Planck, BICEP/Keck and BAO data \cite{BICEP:2021xfz}. } %
    \label{fig:m4}%
\end{figure}

Fig. \ref{fig:m4} depicts the same observables as Fig. \ref{fig:m1.5}, Fig. \ref{fig:m2} and Fig. \ref{fig:m3}  but for $m=4$. The general shape for all plots with $m = 4$ is similar to the plots with smaller $m$ values discussed previously.

Fig. \ref{fig:m4}(a) depicts $r$ vs. $n_\text{s}$ when $m = 4$. The shapes of the relations are similar to the ones for  $m = \frac{3}{2}$, $m = 2$ and $m = 3$. For this $m$ value, the bands are overlapping even more than for the previous ones but the trend that the overlap is bigger for larger $n$ values still holds true. Also, all the bands cover even smaller $r$  than for smaller $m$ values.

Fig. \ref{fig:m4}(b) shows $r$ vs. $\delta$, when $m = 4$. The overall shape is very similar to the previous $m$ values, especially to $m = 3$ in Fig. \ref{fig:m3}(b). A slight difference is that the slopes at small $\delta$ are even smaller, with $r$ values increasing at a slower pace with the increase in $\delta$ values.

In Fig. \ref{fig:m4}(c) $n_\text{s}$ vs. $\delta$ is plotted, when $m = 4$. For this $m$ value, there is a significant overlap present at small $\delta$, where there are more points compared to the smaller $m$ values and where the different models are mostly overlapping. At larger $\delta$, the behaviour is more similar compared to Fig. \ref{fig:m1.5}(c), Fig. \ref{fig:m2}(b) and Fig. \ref{fig:m3}(c) and the decrease in $n_\text{s}$ is again more visible $n \geq 2$ (blue, yellow and red).

In Fig. \ref{fig:m4}(d) $\lambda$ vs. $\delta$ is plotted, when $m = 4$. Once more the general shape is similar to all the previously studied cases, and again at smaller(larger) $\delta$ values $\lambda$ is slightly larger(smaller). In this case, when $\delta \lesssim  50$, $\lambda \lesssim  0.01$ (or even much smaller) for all the considered cases.

\begin{figure}[t]
    \centering
     \subfloat[]{\includegraphics[width=0.5\textwidth]{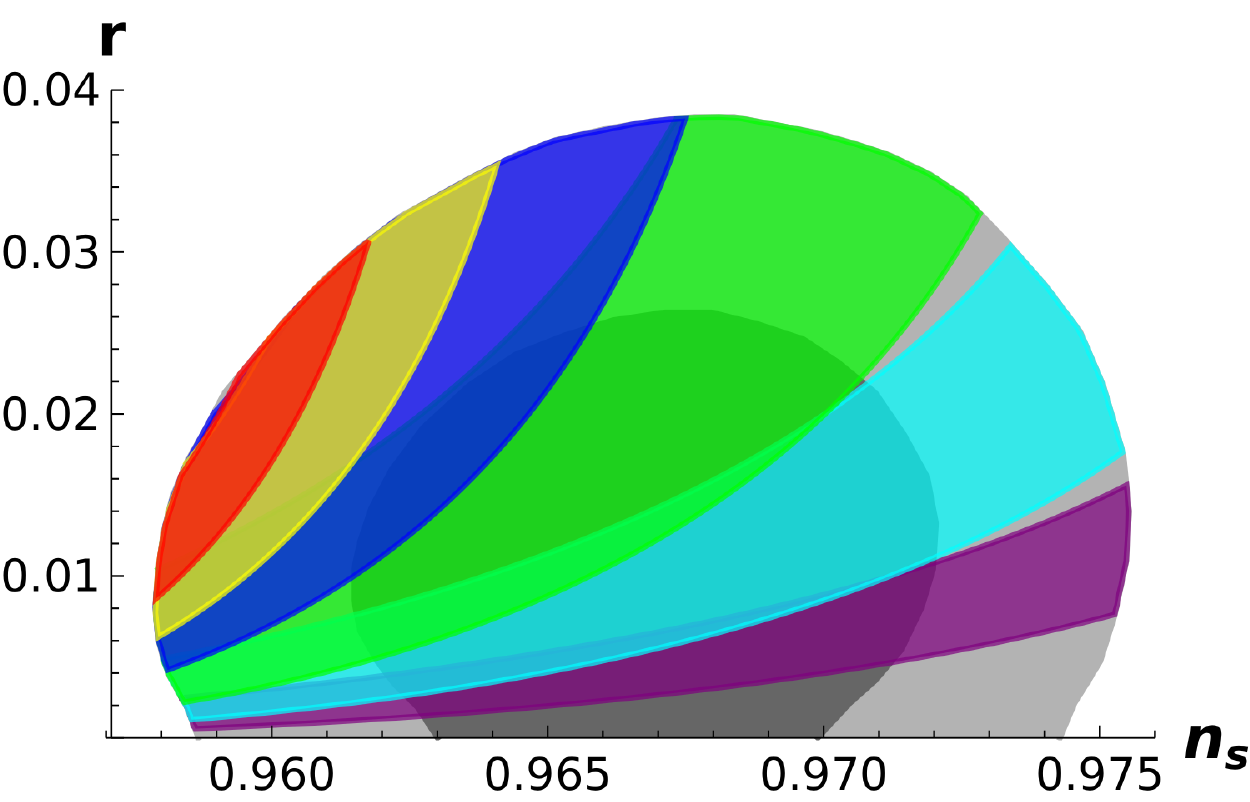}}%
    \subfloat[]{\includegraphics[width=0.5\textwidth]{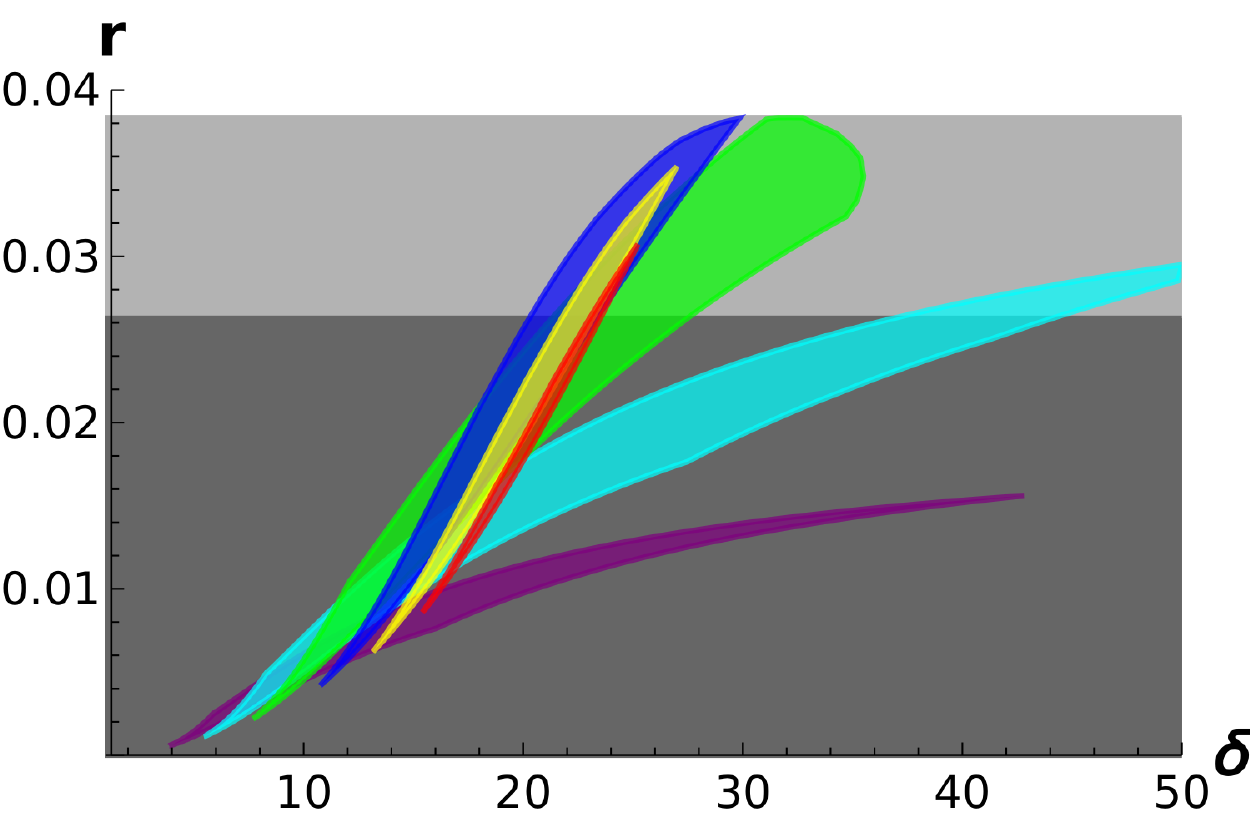}}%

    \subfloat[]{\includegraphics[width=0.5\textwidth]{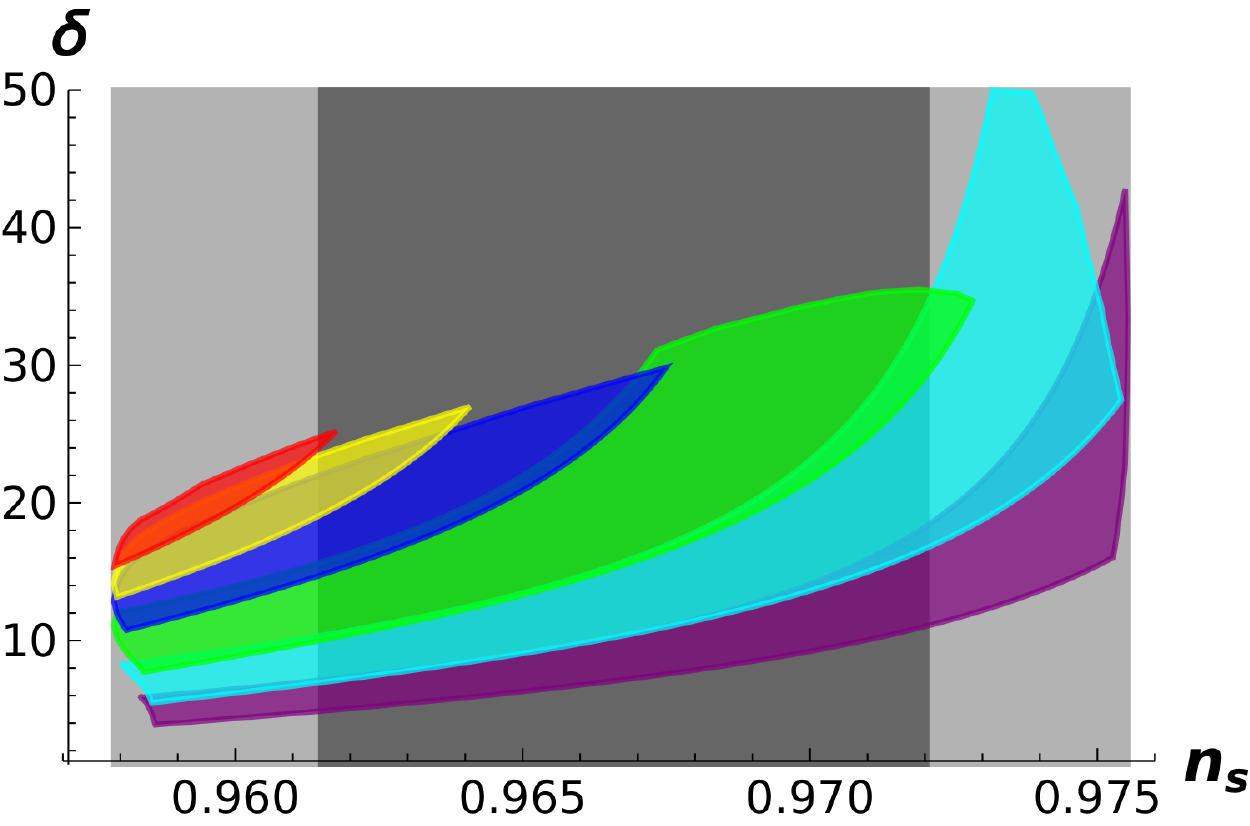}}%
    \subfloat[]{\includegraphics[width=0.5\textwidth]{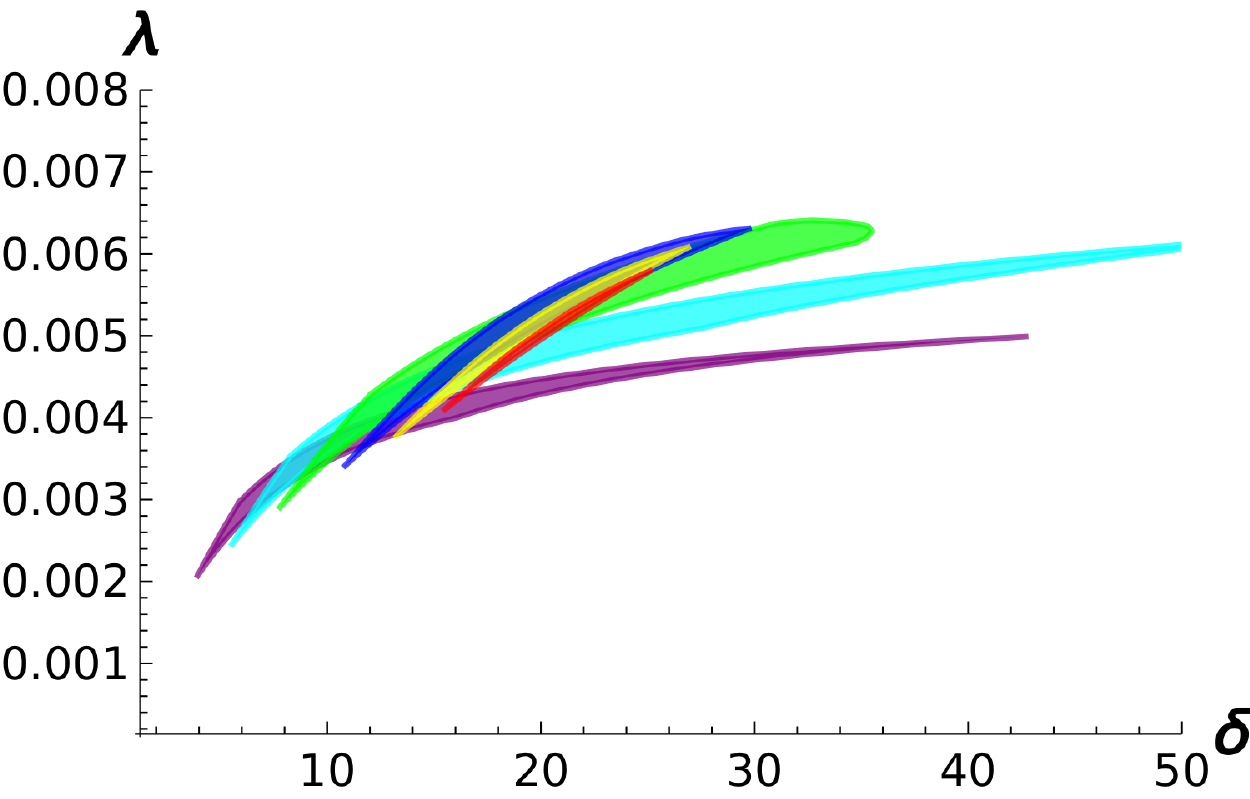}}%

     \caption{$r$ vs. $n_\text{s}$ (a),  $r$ vs. $\delta$ (b),  $\delta$ vs. $n_\text{s}$ (c),  $\lambda$ vs. $\delta$ (d). Filtered plots, where only the points inside the experimental constraints for $r$ and $n_\text{s}$ are kept for $m = 4$ and $N_* \in [50, 60]$. The red areas represent $n = 4$, yellow $n = 3$, blue $n = 2$, green $n = 1$, cyan $n = \frac{1}{2}$ and purple $n = \frac{1}{4}$. With grey are the 1,2$\sigma$ allowed regions from the latest combination of Planck, BICEP/Keck and BAO data \cite{BICEP:2021xfz}. }
    \label{fig:m4_filt}%
\end{figure}


\subsubsection{Filtered results}
The plots in Fig. \ref{fig:m4_filt} follow the same logic as the ones in Fig. \ref{fig:m1.5_filt}, Fig. \ref{fig:m2_filt} and Fig. \ref{fig:m3_filt} (i.e. they are the same as Fig. \ref{fig:m4}, but only points in 2$\sigma$ threshold for $r$ and $n_\text{s}$ constraints are kept and they have been zoomed in). The general shapes are the same as in the previous filtered plots, but the valid areas are larger and this time all the models ($n = \{\frac{1}{4},\frac{1}{2},1, 2,3, 4\}$) are valid.

From Fig. \ref{fig:m4_filt}(a) it can be seen that all of the observed models are valid. The models with $n = 4$ (red) and $n = 3$ (yellow) both have some points inside the 95\% confidence interval. The model with $n = 2$ blue has some points and the models with $n = \frac{1}{4}$ (purple), $n = \frac{1}{2}$ (cyan) and $n = 1$ (green) have many points inside the 68\% confidence interval. All the models taken together cover most of the confidence intervals.

In Fig. \ref{fig:m4_filt}(b) the general shape of the valid regions is roughly the same as for the filtered plots of $m = \frac{3}{2}$, $m = 2$ and $m = 3$. For smaller $\delta$ values, the smaller the value of $n$, the smaller values of $r$ are possible. For larger $\delta$ values, it can be seen that for $n < 2$ the larger the $n$, the larger the corresponding $r$ values. As for $n \geq 2$ applies the opposite, that the larger the $n$, the smaller the corresponding $r$ values. Also, for the models with $n < 2$, the bands increase at a slower pace compared to the ones with $n \geq 2$.

Fig. \ref{fig:m4_filt}(c) depicts $\delta$ vs. $n_{\text{s}}$ in the valid region, with the parameter $m$ being 4. Again, for this $m$, all of the observed $n$ values are valid. The bands are the thickest compared to Fig. \ref{fig:m1.5_filt}, Fig. \ref{fig:m2_filt} and Fig. \ref{fig:m3_filt}. They cover a large area of the confidence intervals.  With the increase in $\delta$ the increase in $n_{\text{s}}$ becomes slower. 

Fig. \ref{fig:m4_filt}(d) depicts $\lambda$ vs. $\delta$  in the valid region, with parameter $m$ being 4. Again, the shapes for different models are roughly the same as in Fig. \ref{fig:m1.5_filt}(d), Fig. \ref{fig:m2_filt}(d) and Fig. \ref{fig:m3_filt}(d). Analogously, the extension of all allowed areas is even larger and the highest valid values for $\lambda$ are slightly smaller than the corresponding ones in the previously studied cases. Considering all $n$'s, the highest valid values for $\lambda$ is still around 0.006 while the smallest is now around 0.002. Now all considered values for $n$ present an allowed region.

\section{Conclusions} \label{sec:Conclusions}
We studied a generalized version of hilltop inflation where the standard hilltop potential has been raised to a power $n$ and we allowed fractional numbers for both the original hilltop power $m$ and the overall exponent $n$. The model was evaluated with different parameters and with the number of e-folds between 50 and 60. The results were compared with experimental constraints coming from the latest combination of Planck, BICEP/Keck and BAO data \cite{BICEP:2021xfz}. We considered $n \in \{ \frac{1}{4}, \frac{1}{2}, 1, 2, 3, 4\}$ and $m \in \{ \frac{3}{2}, 2, 3, 4\}$. 
Since fractional powers are not naturally realized by standard particle physics, we showed in Appendix \ref{sec:appendix} how to generate them from a general scalar-tensor theory. 
The effect of different model parameters on the various observables was investigated. 

At large $\delta$ values, the role of the parameter $m$ is marginal because of the limit in eq. \eqref{eq:monomial:limit}. However, by increasing $m$, the knee of the $r$ vs. $n_s$ regions at $n>1$ moved towards the allowed region, improving the agreement with the experimental constraints. $m=4$ was the only case were compatibility with data was possible for all the considered values for $n$.
The model with $n = 1$ corresponds to the original hilltop model (e.g. \cite{hilltop,Dimopoulos:2020kol,German:2020rpn,Hoffmann:2021vty,Lin:2019fdk,Kallosh:2019jnl} and refs. therein) and agreement with data was somehow possible for all the considered values of $m$, but not always in the whole range of $[50,60]$ $e$-folds. Our results agreed with the previous studies. 
The model with $n = 2$ has also been already studied in the past (e.g. \cite{Hoffmann:2021vty,Kallosh:2019jnl,OLIVE1990307} and refs. therein) and compatibility with the constraints required $m > 2$.  Again our results agreed with the previous studies.
Higher values for $n$ were compatible with data only for $m=4$. 
On the other hand, the models with $n = \frac{1}{4}$ or $n = \frac{1}{2}$ were well within the experimental constraints for any $m$.
Summarizing, agreement with data favors higher $m$ values and lower $n$ ones.

Finally, for all the considered  $(m,n)$ configurations $0.002 \lesssim \lambda \lesssim 0.007$, therefore the energy scale $V_0^{1/4}$ happens to be around the GUT scale. This seems to be a strong indication for an eventual relation between GHI and GUT models.
This might have a relevant impact on model building and ruling in/out inflationary models, especially in light of the increased precision of future experiments (e.g. Simons Observatory \cite{SimonsObservatory:2018koc}, PICO \cite{NASAPICO:2019thw}, CMB-S4 \cite{Abazajian:2019eic} and LITEBIRD \cite{LiteBIRD:2020khw}).

\section*{Note}
This article is based on the BSc thesis of H. G. Lillepalu \cite{Lillepalu}.

\section*{Acknowledgements}
This work was supported by the Estonian Research Council grants MOBTT86, PRG1055 and by the ERDF Centre of Excellence project TK133.

\appendix

\section{Non-minimal action} \label{sec:appendix}
It is challenging to naturally generate potentials with fractional powers. However, it is indeed possible to achieve that via the use of more familiar polynomial functions if non-minimal terms are allowed in the action.
As a proof of concept, we consider the following Jordan frame action in the Palatini\footnote{The choice of Palatini gravity is only dictated by a more immediate field redefinition. It is possible to obtain the same results in the more common framework of metric gravity (see e.g. \cite{Jarv:2016sow,Jarv:2020qqm} and refs. therein).} formulation of gravity
\begin{equation}
S = \!\! \int \!\! d^4x \sqrt{-g^J}\left(-\frac{M_\text{P}^2}{2}f(\varphi)R_J(\Gamma) + k(\varphi)\frac{(\partial \varphi)^2}{2}  - V(\varphi) \right)  \, ,
\label{eq:JframeL}
\end{equation}
where\footnote{ We stress that this is just an example for a more elegant generation of potentials involving fractional powers, fulfilling the purposes of our effective study. The list can be much more longer, given the freedom of choosing the functions $V(\varphi)$, $f(\varphi)$ and $k(\varphi)$.}
\bea
 V(\varphi) &=& a \, \varphi \, f(\varphi)^2 \, , \\
 f(\varphi) &=& 1-\frac{\varphi ^2}{M_1^2} \, , \\
 k(\varphi) &=& \frac{\varphi ^2}{M_2^2} \, .
\eea
We can immediately see that this can only be an effective description (as typical for hilltop models) since we need to require $\varphi < M_1$ in order to avoid repulsive gravity. We can perform the following Weyl transformation
\begin{eqnarray}
\label{eq:gE}
g^E_{\mu \nu} = f(\varphi) \ g^J_{\mu \nu} \, ,
\end{eqnarray}
and move to the Einstein frame
\begin{equation}
S = \!\! \int \!\! d^4x \sqrt{-g^E}\left(-\frac{M_\text{P}^2}{2} R_E(\Gamma) + \frac{(\partial \phi)^2}{2}  - U(\varphi(\phi)) \right) \, ,
\label{eq:JframeE}
\end{equation}
where the Einstein frame scalar potential is
\be
U(\phi) = \frac{V(\varphi(\phi))}{f^{2}(\varphi(\phi))} = a \, \varphi \, (\phi) \, ,
\label{eq:U:general}
\ee
and the canonical normalized scalar is defined by
\begin{equation}
\frac{\partial \phi}{\partial \varphi} = \sqrt{\frac{k}{f}} =   \sqrt{\frac{\varphi ^2}{M_2^2 \left(1-\frac{\varphi ^2}{M_1^2}\right)}} \, .
  \label{eq:dphiP}
\end{equation}
The solution of the field redefinition gives us
\be
\phi= \frac{M_1^2}{M_2} \sqrt{\left(1-\frac{\varphi}{M_1} \right) \left(1+ \frac{\varphi}{M_1} \right)} \, ,
\label{eq:chi}
\ee
which can be inverted into
\be
\varphi=  M_1 \sqrt{1-\frac{M_2^2 \,  \phi ^2}{M_1^4}}  \, .
\label{eq:phi}
\ee
Now inserting it in the potential \eqref{eq:U:general} and choosing $M_1 = \sqrt{M_2 \phi_0}$ and $a=V_0/M_1$ we easily obtain
\be
U(\phi) = V_0 \sqrt{ 1-\frac{\phi ^2}{\phi_0^2}}
\ee
which is exactly the same as eq. \eqref{eqn:potential} with $m=2$ and $n=1/2$.
In a similar fashion also the other combinations of $m$ and $n$ in potential  \eqref{eqn:potential} can be generated.

\section*{Data Availability Statement}
No Data is associated with the manuscript.

\bibliography{GHI}

\end{document}